\documentclass[preprint,11pt]{aastex}
\usepackage{epsfig}

\usepackage{color}

\def\new#1{\color{blue}{#1}\color{black}}

\newcommand{\etal}{et al.~}

\def\msun{\,\rm M_\odot}
\def\Gyr{\,\rm Gyr}
\newcommand{\rvir}{r_{\rm vir}}

\begin{document}

\title{Feedback from galactic stellar bulges and hot gaseous haloes of galaxies }
\author{Shikui Tang, Q. Daniel Wang, Yu Lu, and H. J. Mo}
\affil{Department of Astronomy, University of Massachusetts,
710 North Pleasant Street, Amherst, MA 01003}

\begin{abstract}

We demonstrate that the feedback from stellar bulges can 
\new{in principle
}
play an essential role in shaping the halo gas of galaxies with substantial
bulge components by conducting 1-D hydrodynamical simulations.
The feedback model we consider consists of two distinct 
phases: 1) an early starburst during the bulge formation and 2) 
a subsequent long-lasting mass and energy injection from stellar winds 
of low-mass stars and Type Ia SNe.
An energetic outward blastwave is initiated by the starburst and is
maintained and enhanced by the long-lasting stellar feedback.
For a Milky Way-like galactic bulge, this blastwave heats up the
circum-galactic medium to a scale much beyond the virial radius,
thus the gas accretion into the halo can be completely stopped. 
In addition to that, the long-lasting feedback in the later phase 
powers a galactic bulge wind that is reverse-shocked at a large 
radius in the presence of circum-galactic medium and
hence maintains a hot gaseous halo. As the mass and energy injection
decreases with time, the feedback evolves to a subsonic and quasi-stable 
outflow, which is enough to prevent halo gas from cooling. 
The two phases of the feedback thus re-enforce each-other's impact 
on the gas dynamics.
The simulation results demonstrate that the stellar bulge feedback 
may provide a plausible solution to the long-standing problems 
in understanding the Milky-Way type galaxies, such as the 
``missing stellar feedback'' problem and the ``over-cooling'' problem. 
The central point of the present model is that the conspiracy of the 
two-phase feedback keeps a low density and a high temperature
for the circum-galactic medium so that its X-ray emission is
significantly lowered and the radiative cooling is largely suppressed. 
The simulations also show that the properties of the hot gas in 
the subsonic outflow state depend sensitively on the environment 
and the formation history of the bulge. 
This dependence and variance may explain the large 
dispersion in the X-ray to B-band luminosity ratio of the low 
$L_X/L_B$ elliptical galaxies.  

\end{abstract}

\section{INTRODUCTION}

It is believed that galaxies form in dark matter halos, which grow through
gravitational instability from the density fluctuations seeded in the early
universe. The baryonic matter is then accreted into the dark matter halos
and cools radiatively to assemble galaxies. 
Although the structure and the evolution of dark matter halos are fairly 
well understood with the use of analytic modeling \citep{pres74, bond91, shet01} and 
$N$-body simulations \citep{nfw96, jing00, moor98}, galaxy formation 
still remains a challenging problem as it is a complex process involving 
simultaneous actions of many physical mechanisms. 

One of the long-standing problems in galaxy formation 
is that, in the absence of heating sources, the majority of baryonic matter in 
the Universe is predicted to have cooled into dark matter halos by the 
present time (e.g. \citealt{whit78, whit91}), while in reality
only a small fraction of the baryonic matter associated with galaxies is 
observed in the cold (stars plus cold gas) phase 
(\citealt{Maller04, Mo05, fuk06, Sommer2006}).
This ``over-cooling'' problem implies that feedback must have 
played a very important role in shaping galaxies.
\new{A number of feedback mechanisms have been proposed, which
may significantly affect the formation and evolution of a galaxy,
depending on its halo mass, gas content, and/or star formation
rate. The radiative heating by the extragalactic UV background
may prevent or significantly reduce the accretion of the
intergalactic medium (IGM) onto
a galaxy, which is effective only in small halos with masses
$M< 10^{10}\msun$ at low redshift \citep{quin96, gned00, hoef06}.
The intense UV radiation from a starburst galaxy may further
drive dusty materials into a so-called momentum-driven galactic wind.
In addition, SNe and stellar winds from massive stars may
help to mechanically blow gas out of such galaxies,
and are shown to be effective in galaxies with halo mass less than
$10^{11}\msun$ and become much less effective in massive
halos \citep{MacLow99}. 
For massive galaxies and clusters of galaxies, the AGN
feedback is believed to be more important, particularly in terms of
balancing or reducing the cooling of accreted gas
(e.g. \citealt{wyit03, crot06, Lu07}). The importance of
these feedback mechanisms are, however, much less clear in the
evolution of intermediate-mass galaxies (with halo
masses $\sim 10^{12}\msun$) as are concerned here,
presumably because each of the feedback mechanism can play
some role in affecting the gas accretion and star formation
(e.g., \citealt{Mo02,Toft02,Shankar2006,DO07}). }

Energetically, the ongoing galactic feedback appears to be 
sufficient to balance the cooling. Indeed,
much of the expected mechanical energy from supernovae (SNe) feedback 
is ``missing'' at least from X-ray observations
\citep{David2006,Li06,Li07a,Li07b}.
In general, the observed
X-ray emission consists of several components: 1) relatively bright
X-ray binaries with neutron star and black hole companions ($\sim 10^{36}-
10^{38} {\rm~ergs~s^{-1}}$), which may be detected
individually in nearby galaxies; 2) numerous
cataclysmic variables (CVs) and coronally active binaries (ABs)
($\sim 10^{30} - 10^{33} {\rm~ergs~s^{-1}}$), which are too faint to be
detected individually, but can be constrained based on near-IR
emission \citep{Rev06,Li07a}; 3) diffuse hot gas
heated chiefly by Type Ia SNe. 
The typical observed diffuse X-ray emission only 
accounts for a few percent of the expected SN energy input (e.g., 
\citealt{Li07a, Li07b}), the majority of which is undetected. 
\new{In the low $L_X/L_B$ bulge-dominated galaxies (typically Sa spirals,
S0, and low mass ellipticals) there is little cool gas to hide or
radiate the energy in other wavelength bands. Thus the missing feedback
problem in these galaxies becomes particularly acute.
}
Naturally, one would expect that the missing energy in such galaxies
is gone with a galactic outflow, which could have 
a profound impact on the halo gas and the accretion of the IGM.  
Of course, this impact should evolve with time and depend on the
formation process of the galaxy. For example, if a bulge forms primarily
in a starburst (SB), it can induce an initial
blastwave propagating into the surrounding medium.
The heated circum-galactic medium (CGM\footnote{
  In this paper CGM refers
  loosely as the surrounding medium that has been thermally and/or
  chemically influenced by the galaxy and is located beyond
  the galactic disk and bulge. The IGM is used only for the medium
  that is not chemically enriched during the SB and later feedback.
  Thus the CGM also includes part of the ISM, e.g.,
  the galactic corona or galactic halo gas.})
can then affect the impact of the later feedback from stellar mass
loss and Type Ia supernovae (SNe) as well as from star formation and AGN. 
Therefore, one needs to study the interplay of 
the stellar feedback and the gas accretion of the galaxies to understand
their evolution.

Existing studies have shown the importance of the stellar feedback,
though they are limited by the over-simplified treatments or are
for some specific targets. On the one hand, existing models 
for galactic bulge outflows/winds include little 
consideration for the evolving gravitational 
potential and gaseous circum-galactic environment
(e.g., \citealt{MB1971,WC1983,LM1987,CiotPel1991,David1991, Pel98}).
\new{In these galactic wind models, the stellar feedback mainly includes
the energy input from SNe (primarily type Ia) and mass input from evolving stars.
The specific energy, the ratio of energy input rate to mass input rate,
characterizes the gas state. A galactic wind (i.e., gas leaving the
galaxy supersonically) can be sustained when the specific energy
of the input mass is greater than the gravitational potential
if initially there is only negligible amount of gas inside the galaxy.
}
On the other hand, in simulations of large-scale structure formation, 
the treatment of the stellar feedback is limited by the spatial resolution. 
A more sophisticated model that incorporates
the stellar feedback in an evolving galaxy has been investigated by 
\citet{Bri99} with 1-D hydrodynamic simulations. They found 
that the core-collapse SNe in 
an initial SB can substantially affect the gas dynamics beyond 1\,Mpc. 
But the work was done specifically to explain the hot gas properties
of the luminous elliptical galaxy NGC 4472. In this high-mass galaxy,
later feedback is found to be unimportant, partly because 
the authors adopted a low SN rate (0.03 SNu; one SNu unit denotes one SN per
$\rm 10^{10}\,L_{B\odot}$ per century).
In this case, the gas dynamics is dominated by a cooling flow.

In the present paper, we introduce a stellar feedback model in which 
the feedback is closely associated with the formation history of the galactic bulges. 
In this scenario, we divide the evolution of a bulge into three stages: 
1). At high redshifts, as the host dark matter halo assembles, the central 
galaxy builds up a bulge through frequent major mergers. 
2). As a consequence of the violent mergers, a SB in the bulge is triggered. 
The intensive starburst powers a blastwave to heat and rarefy the CGM.
3). At  low redshift, the galaxy is in a quiescent phase as its host 
halo accretes mass gently. However, as the low-mass bulge stars 
die off from the main sequence, Type Ia SNe provide 
a long-lasting energy input to the CGM and
this gradual feedback maintains a hot gaseous halo. 
This scenario naturally connects the stellar feedback and the general 
formation history of a galaxy. We find that such a feedback can profoundly 
affect the evolution of the model galaxies. In particular,
the overcooling and missing-energy problems can be naturally explained.  
Here we demonstrate the effects of the feedback based
on a 1-D model of galaxies, although limited 2- and 3-D
simulations have also been carried out to check multi-dimensional effects.
\new{Instead of modeling the detailed interaction between feedback sources and
the ISM, we simply assume all the feedback energy is thermalized in the
hot gas of the central halo. }  
Our aim here is to capture qualitatively the feedback effects
on the structure, physical state, and dynamics of the hot gas
around bulge-dominated galaxies on scales greater than galactic bulges/disks.
\new{For the dynamics of hot gas on such a scale,
the detailed mechanisms of the feedback, i.e.,
how the SNe thermalize the ISM in the first place, is not important.
And the efficiency of the thermalization is represented by a free
parameter which in principle should be determined from observations.}

The organization of the paper is as follows. We first describe the main
physical ingredients of our model and numerical methods (\S 2).
We then present the results of a reference galaxy model and its variations
(\S 3). In \S 4, we discuss how the results provide a potential solution
to the overcooling and missing energy problems, as well as the caveats 
of the present model. We summarize our results and conclusions in \S 5.

\section{Galaxy Model and Methodology}

We set up hydrodynamical simulation models to investigate the effects of 
the feedback from a galactic stellar bulge on galaxy formation.
We focus our study on the 
early-type spirals and the typical low$-$$L_X/L_B$ elliptical galaxies 
in fields or in small groups (such as the MW and M31).
Our fidicuial galaxy has an intermediate/massive galactic bulge and 
a dark matter halo with mass $\sim 10^{12}\msun$ at the present time.
In the following, we first describe how such a galaxy assembles mass,
and then introduce the recipe to model
the stellar bulge formation and its feedback. Finally we
describe the numerical codes and methods.
Throughout the paper, we adopt the $\Lambda$CDM cosmology with 
$\Omega_m=0.3$, $\Omega_\Lambda=0.7$ and $H_0=100\,$km/s/Mpc. We 
assume the universal cosmic baryon fraction to be $f_{\rm b}=0.17$.

\subsection{The Growth of Dark Matter Halo \label{sec:halo}}

It has been shown by high resolution 
$N$-body simulations that within the 
virial radius $\rvir$ the density 
profile of a collapsed dark matter halo can be well described by the 
NFW profile \citep{nfw96}, 
\begin{equation}\label{eq:nfw}
\rho_{\rm NFW}(r)={4\rho_0 \over (r/r_s)(1+r/r_s)^2},
\end{equation}
where $r_s$ is a characteristic radius and $\rho_0$ a characteristic 
density of the halo that is related to the collapse time of the halo.
The shape of such a profile is usually characterized by the 
concentration parameter, defined as $c=\rvir/r_{\rm s}$.
Since a virialized halo is defined as a spherical region with a 
mean density that is $\Delta_{\rm vir}$ times the mean density 
of the Universe, the characteristic density can be written as
\begin{equation}\label{eq:hdens}
\rho_0={1 \over 12}\rho_{crit} \Omega_{m} 
\Delta_{\rm vir} {c^3 \over \ln(1+c) - c/(1+c)}\,,
\end{equation}
where $\rho_{\rm crit}$ is the critical density of the Universe.
For the cosmology we adopted here, $\Delta_{\rm vir}=340$ at 
$z=0$ \citep{Bryan98}. For a given cosmology, a 
virial mass, and a given time,  the density profile of a dark matter 
halo is determined by the single parameter, $c$. 
Simulations 
have demonstrated that dark matter halos assemble their mass through
an early fast accretion phase followed by a slow one 
\citep{Wechsler2002,Zhao03a}. 
This two-phase assembly history can be fairly well described by an 
exponential function \citep{Wechsler2002}, 
\begin{equation}\label{eq:mah}
  M(a)=M_0 e^{-2a_c\left(\frac{1}{a}-1\right)},
\end{equation}
where $M_0$ is the halo mass at the present time, 
$a$ is the expansion scale factor
[i.e., $a=(1+z)^{-1}$, $z$ is the redshift],
and $a_c$ is the scale factor at the halo ``formation time''
that separates the two accretion phases. 
Both $N$-body simulations \citep{bull01, Wechsler2002,
Zhao03a, Zhao03b} and analytical model \citep{Lu06} have shown that 
the halo concentration parameter is closely related to the formation time $a_c$.  
For a halo that is still in its fast accretion regime, 
$c\sim 4$; for a halo identified at $a>a_c$, $c\sim 4a/a_c$.

\begin{figure}[htpb]
\begin{center}
\plotone{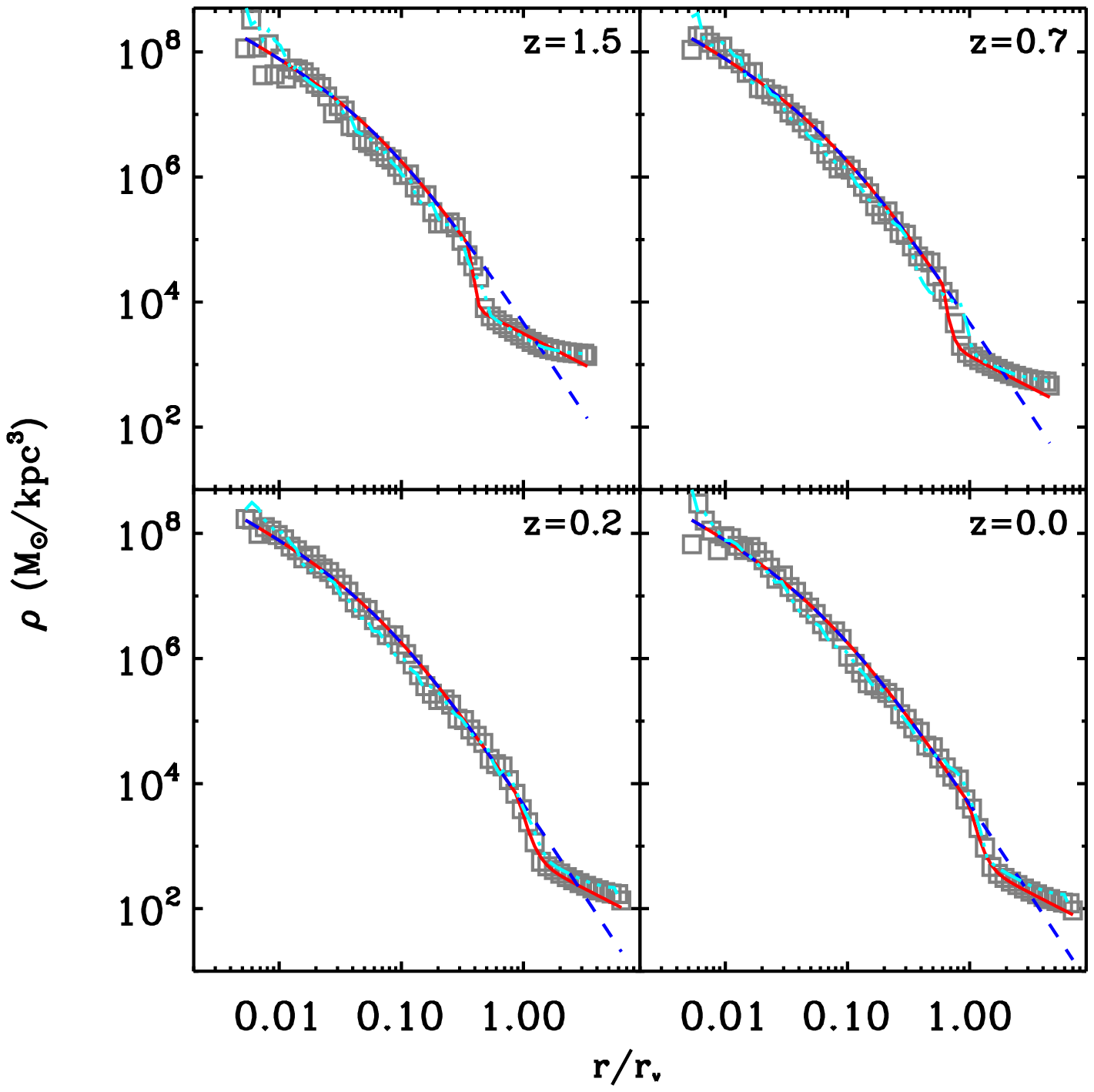}
\caption{\label{Fig:dmprofile}
  Dark matter density profiles of a galaxy of $10^{12}\msun$ at $z=0$
  at four representative redshifts:
  Simulation without stellar feedback (grey squares);
  NFW profile at $z=0$ (dashed blue lines);
  the fitting formula (solid red lines);
  Simulation including stellar feedback (\S2.2) (three-dot-dashed cyan lines).
}
\end{center}
\end{figure}

We make use of the 1-D simulation code developed by \citet{Lu06}
to reproduce the mass distribution of a growing dark halo
embedded in the cosmic density field. 
Following \citet{Lu06} and \citet{Lu07}, we setup the initial 
perturbation at $z_i=100$. The perturbation profile is set up
in such a way that the subsequent accretion history of the halo 
follows exactly the average mass accretion history given by 
Eq.(\ref{eq:mah}). The halo's virial mass is $10^{12}\msun$ at the present 
time. To be consistent with the statistical 
properties of dark halos in cosmological simulations, we choose the formation 
time $a_c=0.3$. The resulting halo has a concentration of about 13,  
which is typical for halos of such a mass. 
The advantage of adopting this approach is that, at any redshift, 
the density profiles both within and outside the current virial radius 
are modeled realistically.  
Fig.~\ref{Fig:dmprofile} shows the density profiles of the simulated halo 
at four different redshifts.
We further find that the evolution of our simulated profile can be
fitted with the following formula:
\begin{equation}\label{eq:dmfit}
  \rho(r,t) = \left\{\begin{array}{l}
      4\rho_0/\left[x(1+x)^2\right],\quad x=r/r_s {\rm~for~} r \leq r_v \\
      \rho(r_v,t)\lbrack\frac{8}{9}\left(\frac{r}{r_v}\right)^{\alpha(t)} +
        \frac{1}{9}\left(\frac{r}{r_v}\right)^{-1} \rbrack,\quad
      {\rm~for~} r > r_v
    \end{array}\right.
\end{equation}
where $t$ is the cosmic time in Gyr;
$r_v$ is the halo virial radius at time $t$
and $r_s$ is the scale radius of the NFW profile at $z$=0;
$\alpha(t)=-46+9.4t-0.58t^2$ is a time-dependent index obtained 
from fitting the simulated profiles at $r \gtrsim r_v$. 
This formula (illustrated as the solid red lines in
Fig.~\ref{Fig:dmprofile}) characterizes the simulated profiles
to an accuracy of $5\%$ in the interested region.
We adopt this fitting formula to simplify the gravity calculation of
the dark matter in Eulerian-based hydro-dynamical simulations.
Although the deposition of the cold gas in the halo center may make the 
the halo contract \citep{blum86, choi06}, our test with
the same 1-D simulation code but including the gravity of the 
cool gas shows that the change of the dark matter 
distribution is only in the central several kpc, 
and that our main results on the halo gas properties 
are not expected to be affected significantly by such 
contraction.

\begin{deluxetable}{lr}
  \tabletypesize{\footnotesize}
  \tablecaption{\label{tab:para} Model galaxy parameters}
  \tablewidth{0pt}
  \tablehead{
  \colhead{Parameter} &
  \colhead{Value}
}
\startdata
Halo mass  $M_0 {\rm~(M_\odot)}$&$10^{12}$\\
SB time $t_{sb}$ (Gyr) ($z=1.8$) &2.5\\
SB gas mass $M_{sb} {\rm~(M_\odot)}$&$5 \times 10^{10}$\\
SB energy deposit region $R_{sb}$ (kpc) & $50$\\
Bulge stellar mass $M_*{\rm~(M_\odot)}$&$3 \times 10^{10}$\\
Bulge scale $r_b$ (kpc) & $ 0.7 $\\
Age of galactic bulge $t_0$ (Gyr)& 10 \\
Bulge blue band luminosity ($L_{B,\odot}$) & $5\times 10^{9}$ 
\enddata
\end{deluxetable}

\subsection{Stellar Bulge Feedbacks}

We approximate the stellar feedback into two phases: (1) a blastwave generated 
by the initial SB that forms the stellar bulge and 
(2) a gradual injection of mass and energy through stellar mass-loss
and Type Ia SNe in the bulge.
Additional feedback may arise from 
the galactic disk, if present (although cool gas there
may consume much of the input energy), and from AGNs. We do not deal with these
complications here, which may be incorporated into the above two 
phases approximately within the uncertainties in our adopted feedback parameters. 
The key parameters are the initial SB energy feedback $E_{SB}$ and
the evolution of the Type Ia SN rate and mass loss rate.

\subsubsection{Starburst feedback}

We assume that the bulk of bulge stars forms in a single SB at the time
$t_{\rm sb}$ right after the fast-accretion regime of the dark matter 
halo, when the inner halo is well established.
In this regime the gas cools in a time-scale much less than
the halo dynamical time-scale; dense gas clouds are thus expected to 
form rapidly, leading to the SB \citep{Mo04}.
The SB converts a fraction of the cooled mass $M_{\rm b}$
into bulge stars with a total mass $M_*$
comparable to that observed in our MW or M\,31.
The adopted values are listed in Table \ref{tab:para}.

We parameterize the SB energy feedback with an estimate of the mechanical energy
input from core-collapsed SNe  ($M \gtrsim 8 M_\odot$).
This estimate depends on the
initial mass function (IMF) of stars formed in the SB.
Assuming the commonly used Salpeter IMF over the mass range of 0.08-100
M$_\odot$, the number of stars with individual masses greater than
8$\msun$ is $N_{\rm >8} = 6.8\times 10^{-3} M_*$.
The total energy deposited into the large-scale surrounding is
\begin{equation}\label{eq:enSNeII}
   E_{\rm SB}=\eta E_{\rm sn}N_{\rm >8} =
   6.81\times 10^{-3} \eta M_* E_{\rm sn},
\end{equation}
where $E_{\rm sn}$ (=$10^{51}$ ergs) is the energy released by each SN.
The parameter $\eta$ characterizes the ``efficiency'' of SB feedback, 
which in principle accounts for the uncertainties in the assumed
IMF, $E_{\rm sn}$, possible energy release from other sources (e.g., AGN), and
the radiative dissipation within the SB region.
The radiative dissipation, for example, depends on the unknown details
of the interaction between the SB and the ISM
(e.g., its porosity; \citealt{Silk03}). 
And a probably more realistic IMF with a low mass turn-over
(e.g., \citealt{MS79,Kroupa01})
can give about 10\% higher energy input.
\new{In addition, the UV radiation released by massive stars
at the main sequence stage is also a considerable energy source for
the feedback, and the total radiation energy is comparable to or even
larger than the mechanic energy output from the SN explosion of
the massive stars \citep{Schaerer03}.
The UV radiation may contribute to the ionization of the ambient medium,
but is mostly re-radiated away at lower frequencies. A fraction of the
radiation may escape from star forming regions, but is expected to
contribute little to the energy balance in the circum-galactic medium,
especially its hot component with $T \gtrsim 10^6$ K. We have assumed that
the medium is ionized initially, which could be due to the combination
of the extragalactic background and the escaped ionizing radiation
from the model galaxy. Probably the most important dynamic effect of
the UV radiation is its momentum inserted on dust, which may help
to drive the expansion of the gas away from the galaxy. This
contribution is absorbed into the efficiency parameter $\eta$, which is
treated as a free parameter in our model and can only
be calibrated and determined by observations. }

\subsubsection{Gradual feedback}

Following the core-collapsed SNe of massive stars is the long-lasting
but relatively gentle feedback from stars of lower masses.
Within the uncertainties in stellar evolution models
the energy input rate from Type Ia SNe can be approximated as
(e.g., \citealt{CiotPel1991,Bri99})
\begin{equation}\label{eq:snl}
  L_{SNIa}(t) = E_{\rm sn} \left(\frac{L_B}{10^{10}}\right)
  \frac{R_0}{100} \left(\frac{t}{t_0}\right)^{-p} {\rm erg\,yr^{-1}},
\end{equation}
where $t$ is the age of stellar population,
and $L_B$ is the blue band luminosity of the stellar
population at the present age $t_0$ (Table 1), which can be determined
from a stellar evolutionary model (e.g., \citealt{Mar05}).
The Type Ia SN starts $\sim 0.1$\,Gyr \citep{Greg05,Man06,Aub07}
after the starburst.
Since the stellar properties in galactic bulges
are similar to those in ellipticals, we set $R_0$=0.12 (SNu)
which is the average local SN Ia rate for E/S0-type galaxies
(\citealt[p467]{Cap99,Ar99}).Depending on the choice of 
models of SNIa progenitors \citep{Greg05}, $p\sim1.0-1.5$.


Most of the stellar mass loss occurs at early time through the relatively
fast evolution of intermediate-mass stars.
\new{The mass loss rate of a stellar population is given by the product of the
stellar death rate and the average mass that is lost by each dying star. }
We use an updated calculation obtained from \citet[Fig.~22]{Mar05},
which directly relates mass loss to the total initial mass of a stellar population.
The cumulative stellar mass loss of a normalized stellar population with initial
mass of 1\,$\msun$ can be fitted by the following formula:
\begin{equation}\label{eq:fitmlr}
  M(<t) = 0.349 - 0.122 t ^{-0.3}
  + 0.0737 e^{-t/0.158},
\end{equation}
which matches the numerical result to an accuracy of
1\% from 0.1\,Gyr to 15\,Gyr. An alternative formula introduced
in \citet{CiotPel1991} gives the same shape as that of Eq.~(\ref{eq:fitmlr})
for $t>1$\,Gyr.

It is worth comparing the relative energetics of
the stellar feedback during the SB and the later gradual phase. 
There is $6.8\times 10^{-3}$ SNe II per solar mass
for a standard Salpeter IMF. For a general power law decay of the
SNe Ia rate, the number of SNe Ia per solar mass is
$ 10^{-3} R_0 (M/L_B)^{-1}  n_{_{Ia}} $, where $(M/L_B)$
is the mass to light ratio, and 
\begin{equation}\label{eq:numsneIa} 
  n_{_{Ia}} = \int^{t_0}_{t_{em}}
  \left(\frac{t}{t_{0}}\right)^{-p}dt.
\end{equation}
With the reasonable ranges of SNe Ia emerging time $t_{em}$
(e.g, 0.1-0.5\Gyr) and $p$ (e.g, 1.0-1.4), $n_{_{Ia}} \simeq 100$
with an uncertainty about a factor of two.
The $M/L_B$ for a single stellar population with standard Salpeter IMF
is about 7 at 10 Gyr \citep{Mar98}.
Thus the number of SNe Ia per solar mass
is $\sim 1.7\times 10^{-3} \left(\frac{R_0}{0.12}\right)
\left(\frac{7}{M/L_B}\right)\left(\frac{n_{_{Ia}}}{100}\right)$.
SB feedback in principle provides much more energy
in a short time than the gradual feedback does in the life time
of a galactic bulge.
Note that the time in Eqs.~(6)-(8) is based on ``stellar clock''
and modeled independent of cosmology.
We assume that the ``stellar clock'' is consistent
with the concordance cosmology model
(where $H_0=70\rm km\,s^{-1}\,Mpc^{-1}$),
and scale the time unit in Eqs.~(\ref{eq:snl}) and (\ref{eq:fitmlr})
accordingly in our simulations.

\subsection{Methods for Gas Evolution Simulations}

We adopt the 1-D Lagrangian hydrodynamical code \citep{Lu07}
to simulate the accretion of the gas and dark matter
during the fast assembly regime.
As introduced in \S\ref{sec:halo}, we use 1000 and 100,000 shells
to represent gas and dark matter, respectively.
Before the mass collapses into the virial radius the two components
follow each other. When a gas shell is shocked roughly at the virial radius,
the gas is decoupled from the dark matter flow
and is heated up to the virial temperature (see also \citealt{Bir03}).
The gas then cools radiatively.
When the temperature falls below $2\times 10^4$\,K, the gas shell is
considered to be cold and is dropped out from the hot phase. We re-distribute 
the cold gas 
according to a Hernquist profile \citep{Her90} with the scale radius of 0.7\,kpc.
We terminate the 1-D Lagrangian simulation at $z=1.8$ when the fast-accretion 
regime is over. 
This simulation provides both the mass of the dropped-out gas ($M_b$) and
the density, temperature, and velocity profiles of the remaining hot gas.
The output of gas properties from the Lagrangian code is adopted
as the initial condition for our feedback simulations.

After the fast assembly epoch, the bulk of the galactic
bulge has been built up and the SB is triggered.
The mass and the SB feedback energy are deposited instantaneously and
uniformly in form of thermal energy within the radius $R_{SB} = 50$ kpc,
which is about the virial radius of the halo at the assumed SB epoch
(Table~\ref{tab:para}).
The included gas is assumed to be well mixed with the ejecta of the
SNe and is thus considered to be SB-enriched.
We have experimented various methods to deposit the SB energy
(e.g., kinetic vs. thermal energy as well as radius) and
find that they don't make significant differences as long as no
additional dissipation is allowed before the deposited energy emerges
into the large-scale surrounding.
\new{As we focus on the potential impact of the SB on the ambient
hot gas over much larger physical and time scales, the detailed
evolution of SB itself or its immediate interaction with the
cold gas in the bulge is not important (e.g.,  \citealt{ss00, Fu2004}).
} 

With the feedback included, we conduct our hydrodynamical simulations
using the widely-used FLASH code \citep{Fry00}.
This Eulerian based code makes it easier to
handle the position-wise mass and energy input.
The mass and energy input source terms, specified by Eq. (6) and (7),
are handled by the operator split method in FLASH,
\new{i.e., at each time step it first solves the Eulerian equations
at each pixel without mass and energy input source terms,
then explicitly updates the values at each pixel to account for
the corresponding mass input and net energy input (heating plus cooling)
during that time step.
}
The gravity force includes the dark matter halo with a time-dependent
density profile described by Eq.~(\ref{eq:dmfit}) and a fixed Hernquist
profile for the stellar bulge with a total mass of $M_{sb}$
(Table~\ref{tab:para}). The gravity is dominated by the bulge
stellar mass at small radii and by dark matter at large radii.
The gas self-gravity is thus ignored. 

The radiative cooling of the gas depends on its metallicity as well
as its local density and temperature in the simulations.
We assign representative metallicities of 0.01, 0.1, and 1 solar to
the IGM, the SB-enriched gas, and the Type Ia SN-enriched stellar material.
A 0.1 solar abundance of the SB-enriched gas corresponds to
roughly a yield of 0.02$\,M_\odot$ metals per solar mass of the SB.
The metallicity-dependent cooling function is
adopted from \citet{SD93}. Radiative cooling is switched off for
gas of temperature below $2\times10^4$\,K, which could 
be maintained by the radiation heating.
The different metallicities are also used to track the individual gas
components. 

We have chosen a set of models to illustrate how the dynamics and
observational properties of the hot gas depend on the
SB energy output efficiency ($\eta$),
the gradual stellar feedback evolution,
and the treatment of cold gas.
The model parameters are summarized in Table \ref{tab:feedback}.
The reference model (Model Ref.) adopts the canonical values,
which sets the reference for comparison with other models.
We demonstrate the effects of $\eta$ in Model VA,
the varied gradual mass loading in Model VB, 
and the varied Type Ia SN feedback evolution in Model VC and VD.
The effect of cool and hot gas coupling on the gas dynamics
is examined in model VE.

Our simulation radius is 2\,Mpc around a model galaxy, including 
all the region that the feedback can influence. Because gas at this radius is 
still participating in the Hubble expansion, 
an ``outflow'' (zero-gradient) boundary condition is adopted, and the standard
``reflection'' boundary condition is used at the center.

\section{Results}

In this section, we first focus on the fiducial model 
and then check the variance due to the changes in the model parameters.
We further give the immediate predictions of our model on the halo gas fraction 
and X-ray brightness. 

\subsection{Dynamical Structures and Evolution}

\subsubsection{Reference Model}

Fig.~\ref{Fig:mwb4a} illustrates the gas properties
of the reference model galaxy at three representative redshifts. 
Soon after the SB (e.g., $z$=1.4; dot-dashed line), a blastwave 
(corresponding to the jump at $r \sim 10^{2.5}$ kpc in all the four panels) is 
driven into the accretion flow of the galaxy.
This accretion flow is most apparent in the radial velocity panel (i.e., 
the negative velocity just in front of the blastwave). 
At larger radii, the gas is still trailing the Hubble expansion. 
Later, the blastwave, though weakening, catches up 
the trailing gas (i.e., no negative velocity at $z = 0.5$ and 0.0). 
The blastwave heats the gas to a temperature of $\gtrsim 10^6$ K, 
over a region extending well beyond the virial radius at $z \sim 1.4$
($R_{vir}\sim 80$ kpc) and even at $z =0.0$ ($R_{vir}\sim 200$ kpc). Consequently, 
{\sl after this pre-heating by the SB-induced blastwave, no virial shock
forms.}

The SB heating also allows for the
transportation of the later stellar feedback away from the galactic bulge.
Without the SB heating and the resultant rarefaction of the surrounding gas,
the energy from the Type Ia SNe would be radiated away around the bulge and
the gradual feedback could never launch a galactic bulge wind.
With the low density of the ambient gas due to the expansion,
the gradual stellar feedback develops into a galactic bulge wind, 
at least in early time when the SN rate is high. This wind reaches 
a velocity up to $\sim 10^3 {\rm~km~s^{-1}}$ at $z=1.4$ and $\sim 800
 {\rm~km~s^{-1}}$ at $z=0.0$ as the rate decreases. The wind is 
reverse-shocked at a quite large radius (the innermost jumps of the 
profiles in Fig.~\ref{Fig:mwb4a}).
The reverse-shock first moves outward and then starts to move back toward
the center as the Type Ia SN rate declines with time.
The reverse-shocked wind material has a temperature of $\sim 10^7$ K.
This temperature is determined by the specific energy of the feedback, 
minus the galaxy's gravitational potential that the wind needs to climb through. 
The radiative cooling of the wind is negligible
(see Table 2 and later discussion in \S4.2).
Therefore, the SB heating enables the impact of the Type Ia SN-driven 
feedback over global scales around the galaxy.

In return, the long-lasting Type Ia SN-driven feedback helps to maintain
and even to enhance the low-density hot environment produced by the SB. First,
the inauguration of the Type Ia SN-driven feedback about 
0.1 Gyr after the SB generates a forward shock,
which can be seen in Figure~\ref{Fig:mwb4a} (bottom panel) as a small
velocity jump just behind the SB blastwave. This shock
propagates fast in the SB-heated gas with little radiative energy loss
\citep{Tang05} and quickly overtakes the SB blastwave.
Secondly, the Type Ia SN-driven stellar feedback prevents the SB-heated 
gas from falling back toward the galactic center to
form a cooling flow, which would generate too large an X-ray luminosity to 
be consistent with observations.

Sandwiched between the shocked bulge-wind material and the blastwave-heated 
IGM is the SB-enriched gas 
(e.g., marked as the gray region for the $z = 1.4$ density profile). 
The two contact-discontinuities around the gas are apparent in the
density and temperature profiles of Fig.~\ref{Fig:mwb4a}.
The SB-enriched gas has a relatively high density and a low temperature;
and the gas has a cooling time scale of $\sim 1$\,Gyr
even with the assumed moderate metallicity.
The gas can cool to the lowest temperature 
allowed ($2\times 10^4$\,K in our simulation) first, as represented by 
spike-like features in the density profiles at lower redshifts. 
This collapsed dense shell is subject to various hydrodynamical 
instabilities (if not prevented by the presence of magnetic field);
the fragmented cool gas may be decoupled from the hot gas flow. 


\begin{deluxetable}{lrrrccccc}
  \tabletypesize{\footnotesize}
  \tablecaption{\label{tab:feedback} Models of stellar feedback}
  \tablewidth{0pt}
  \tablehead{
   \colhead{Model} & \colhead{$R_0$} & \colhead{$\eta$}  & 
   \colhead{$p$} & \multicolumn{2}{l}{$^{a}L_X\,(10^{36}\rm ergs~s^{-1})$ } & & 
   \multicolumn{2}{c}{$f_b$}   \\
    \cline{5-6} \cline{8-9} \vspace*{-10pt} \\ 
  \colhead{} & \colhead{(SNu)}  &   &  & ($r\!\leq\!2\rm kpc$) & ($r\leq\!\rvir$) & &
  ($r\!\leq\!\rvir$) & ($r\!\leq\!2\rvir$)
  }
\startdata
Ref.   & 0.12  & 0.5 & 1.4 & 0.25  & 0.45  & & 0.045 & 0.040\\ \hline
VA     & 0.12  & 0.25 & 1.4 & 0.25 & 18.6  & & 0.047 & 0.111 \\
VB$^b$ & 0.12  & 0.5  & 1.4  &  5.7 & 50.0  & & 0.047 & 0.082\\ 
VC     & 0.06$^c$& 0.5  & 1.6  & 1.2  & 2.8 & & 0.046 & 0.041 \\
VD     & 0.35$^c$& 0.5  & 1.0  & 0.038 & 0.064 & & 0.045 & 0.039 \\ \hline
VE$^d$ & 0.06  & 0.25 & 1.4  & 3.8  &  1673 & & 0.163 & 0.168
\enddata
\tablenotetext{a}{$L_X$ is calculated in the 0.3-2.0\,keV band, assuming solar abundance.
  $r$ is the galactocenter radius in units of kpc.}
\tablenotetext{b}{the mass loss rate determined by 
  Eq.~(\ref{eq:fitmlr}) is doubled.}
\tablenotetext{c}{$R_0$ is normalized to give the same total energy input
  as the reference model.}
\tablenotetext{d}{The ``dropout'' recipe is applied (\S\ref{sdropout}).}
\end{deluxetable}

\subsubsection{Dependence on the SB feedback efficiency}

The most uncertain parameter in our reference model is $\eta$,
which determines the magnitude of the energy feedback
during the SB phase.
In Model VA we reduce the SB feedback efficiency to $\eta=0.25$.
The results are compared with those from the reference model in
Fig.~\ref{Fig:mwb4cmp} at $z=0$. 
Naturally, a smaller  $\eta$ value gives a slower and weaker SB blastwave,
hence a relatively higher density and faster cooling of the interior gas.
The velocity becomes negative around virial radius and the material 
blown away earlier is being re-accreted into the halo. 

This relatively high density circum-galactic environment affects the dynamics
of the Type Ia SN-driven feedback, which is confined to a smaller region 
(Fig.~\ref{Fig:mwb4cmp}). An interesting feature 
of this model is the higher temperature of the 
reverse shocked bulge wind material, which has not lost much of its energy
in climbing out the gravitational potential of the galaxy.
The reverse shock is moving inward and is only about 20\,kpc away
from the center of the galaxy at $z=0$. When it eventually reaches the center,
the bulge wind will be quenched. After that, a cooling flow of previously
reverse-shocked bulge wind material will form.
The cooling flow, together with the infall of the already cooled
SB-enriched shell may trigger a new SB in the bulge.
In general, the smaller $\eta$ is, the earlier the 
reverse shock front moves back to the center.

\subsubsection{Dependence on the gradual feedback}

A reduced Type Ia SN energy input rate (e.g., due to a smaller 
$R_0$ and/or $E_{SN}$ than the assumed) has a similar effect as that of
a reduced SB feedback on the galactic gas dynamics:
the decrease of energy input cannot support a galactic wind or
an outflow and even results in an inflow at later time. 
We test a model with half of the rate adopted
in the reference model, and find that the reverse shock and the cooled
SB-enriched dense shell moved back to the center about 2\,Gyr ago.
The resultant flow has an artificially cold core with a high
inner density, similar to the figures shown in \citet[fig.~2]{CiotPel1991}
and \citet[Fig.~3]{Pel98}, which consumes nearly all the energy
feedback locally near the bulge center in the 1-D single phase modeling.
Such a cold core should undergo a nuclear starburst providing additional feedback.
We do not intend to address such complicated situations here.

A reduction in the mass loss rate has the opposite effect. With the
same energy input, but a smaller mass loss rate, the specific heating
of stellar ejecta would increase, helping to 
\new{launch a stronger bulge wind which has a lower gas density. A
reduced mass loss rate would also decrease the diffuse X-ray emission
since it is proportional to the density square.
}
In reality the mass loss rate is most likely larger than the
value calculated from Eq.~(\ref{eq:fitmlr}),
which is based on a single burst of star formation scenario.
\new{In addition to the mass loss from evolving stars, mass-loading from
other sources may also be important, though very uncertain. Such
sources include the evaporation of cool gas clumps (e.g., cloudlets 
formed from the leftover of the SB, cold inflow streams from the IGM,
etc.).
}
In Model VB we double the stellar
mass loss rate given by Eq.~(\ref{eq:fitmlr}).
In this model the inner bulge wind region has shrunk to a radius less
than 20\,kpc at $z=0$ (dotted grey line in Fig.~\ref{Fig:cmpp}).
Inside the shocked bulge wind region the 
density is about twice that of the reference model,
and the density profile is leveling off outside. 
An even larger mass loss rate can quench the inner
bulge wind quickly, inducing an inflow by $z=0$.

The gas dynamics is also affected by how the Type Ia SN rate
evolves (e.g, the choice of $p$).
The inner bulge wind can be established more easily initially
in a model with faster evolutionary trend (i.e, with a larger $p$).
To examine this effect we adopt $p=1.6$ in model VC,
and adjust $R_0$ to ensure that this model has
the same total amount of SNe Ia energy input as the reference model.
The profiles at $z=0$ are shown in Fig.~\ref{Fig:cmpp} (dashed red lines).
In this model more SNe Ia explode at early time 
so that the initial blastwave is stronger and can drive 
the outermost shock front further away than the reference model does. 
But at $z=0$ the resultant bulge wind region
is smaller and the temperature is lower because less
energy is available to maintain the galactic flow at the late stage.
A similar test is done for $p=1.0$ (Table 2 model VD).
Though initially the blastwave is weaker and the outermost 
shock front has a smaller galactocentric radius,
this model maintains a much stronger bulge wind at $z=0$
because it has a higher Type Ia rate at the present time.

\subsubsection{Dependence on the treatment of cooled gas \label{sdropout}}

The effect of the collapsed dense shell on the hot gas flow depends on
the coupling of the cooled and hot gases. 
While the physical processes will be studied in our 2-D modeling (Tang 
\& Wang 2008, in preparation), here we examine the effect in an extreme 
case by adding a simple ``dropout'' recipe in our simulation to decouple
the cooled gas from the hot gas flow. When a gas pixel cools below the
minimum temperature allowed, we dropout 99\% of its mass and 
increase its temperature accordingly to keep the pressure unchanged. 
Using this recipe, the dropout does not generate numerical perturbation
into the simulation and conserves the energy in the flow.
Simulations with or without the dropout
can thus be used to demonstrate the full range of possible dynamic effect.
The dropout in general has little effect on the gas dynamics, if the
cooled gas shell is located at a large radius. This is the case for the
reference model (Fig.~\ref{Fig:cmpdo}). With the dropout, the flow just
becomes a little ``lighter''. But because the gravitational potential
is shallow at the large radius, the effect is small.

On the other hand, if the cold dense shell gets close to
or actually falls toward the center,
the effect on the gas flow can be substantial. The shell imposes a weight
that the enclosed hot gas needs to hold, which can result in increased 
radiative cooling and eventually quench the bulge outflow.
As mentioned previously, for models with a reduced $R_0$ or $\eta$
the dense shell moves back to the center and forms an artificially
cold core before $z=0$. When this weight is removed through
the dropout, the bulge outflow becomes faster and the radiative cooling
becomes slower. The gas evolution of the Model VE
with the ``dropout'' enabled is shown in Fig.~\ref{Fig:stado}.
In this model, the stellar ejecta is confined in a region less than 100\,kpc,
the initial inner bulge wind is quickly quenched and turns into a
subsonic outflow.
The outflow is stagnated around half of the virial radius
(i.e., stagnation is defined as the outflow velocity becomes zero),
and an accretion flow is built up near the virial radius.
Thus a density enhanced region is formed
between half of the virial radius and the virial shock.
Effective cooling first occurs in SB-enriched material because 
it has a relatively high density and low temperature
(see also Fig.~\ref{Fig:mwb4a}).
Most of the dropout occurs at radius larger than 100\,kpc in this simulation.
By dropping out the cooler gas, the galactic flow
can stay in a quasi-static state for quite a long time,
i.e., as for the specific case (model VE) the density and temperature
profiles are hardly changed for half of the Hubble time.
The outermost shock front (actually powered by SNe Ia which overtakes
the blastwave of SB) is still riding on the turnaround 
Hubble flow at $z=0$.
Note that a global galactic wind would still be maintained by the gradual
stellar feedback if it were started from a {\it gas free} initial condition
(Fig.~\ref{Fig:stado}).

\subsection{Baryonic content and X-ray Emission}

The stellar feedback results in a baryonic mass deficit within the virial radius.
The radial variation of the local baryonic fraction (the baryon to total
gravitational mass at each galactocentric radius)
of the reference model is shown in the left panel of 
Fig.~\ref{Fig:mf}, and the corresponding accumulative baryonic
fraction (the total baryonic mass to the total gravitational mass within that radius)
is shown in the middle panel. The dash-dot blue line
denotes the radial variation at the starburst time $t_{sb}$.
The variation of the local baryonic fraction at this time follows the
fluctuation of the simulated dark matter profile given by the
1-D Lagrangian code (see Fig.~\ref{Fig:dmprofile}).
The cumulative fraction shows that the baryonic mass fraction decreases
nearly monotonically and reaches the universal value around the virial radius.
At $z$=0, the baryon fraction
drops down to about 5\% near the virial radius and reaches the minimum
value at the contact-discontinuity between the materials from the two
feedback phases. The baryon fraction then increases
outward and finally reaches the universal fraction at the forward blastwave
(located at a distance about 6 times the virial radius).
This increase is largely due to the SB-enriched material and
the CGM swept up to a large radius by the galactic wind.
Within the virial radius, the hot gas contribution is negligible,
as shown by the dash red line (the inner part has been 
scaled up by a factor of $10^3$ in the plot for easy visualization).

The amount of baryonic mass within the virial radius depends on feedback parameters,
such as the strength of the SB and the gradual stellar feedback evolution.
The cumulative baryonic mass fraction of Model VA, for example, is shown in
the right panel of Figure~\ref{Fig:mf}. With the reduced SB efficiency,
the baryonic mass fraction at $z=0$ is higher around the virial radius
than that of the reference model.
In Table 2 we list the baryonic mass fractions for all the models
within one and two virial radii, respectively.
For models with strong SB feedback (i.e., Models Ref., VC, and VD),
the baryonic fractions within one virial radius are about 5\%,
and the fractions are even lower within two virial radius,
which demonstrates that a large fraction of the baryonic mass
is expelled out to a region beyond two virial radius.
In Model VE the baryonic fraction is slightly less than the universal
value because of its reduced stellar feedbacks.

The X-ray luminosities are listed in Table 2 for all the models.
\new{We assume that the hot gas is in collisional ionization equilibrium
and calculate the luminosities based on the 0.3-2.0 keV emissivity
as a function of temperature, which are extracted from the {\bf xspec}
software for three different metallicity we adopted.
}
The luminosities within 2\,kpc characterize
the diffuse X-ray emission from the inner bulge wind.
The X-ray luminosity of the reference model in the inner region
is $\sim 2.5\times10^{35}\rm~ergs~s^{-1}$,
and in Model VE the X-ray luminosity is
$\sim 4\times10^{36}\rm~ergs~s^{-1}$.
The X-ray luminosities within the virial radius
vary dramatically with respect to different models.
In the reference model it is only $\sim 4.5\times10^{35}\rm~ergs~s^{-1}$,
whereas in Model VE the luminosity is as high as 
$\sim 2\times10^{39}\rm~ergs~s^{-1}$.

\section{Discussion}

We now discuss the implications of the above results in connection
to the issues mentioned in \S 1 and the additional work
that is needed to further the study of the feedback scenario 
proposed here. 

\subsection{The Over-cooling Problem}

\new{Our simulations demonstrate that the total accreted baryonic mass can
be largely reduced when the stellar feedback from both an initial SB
and a later input from evolving stars is implemented self-consistently
in the halo (galaxy) formation scenario.
}
Except in Model VE, the baryonic mass fractions are only about 5\%
within virial radius, consistent with an inferred low baryonic fraction
in the Milky Way (roughly about 6\%, 
e.g., \citealt{Mc07, fuk06, Sommer2006, Maller04}).
In our reference model, the radiative cooling after the SB is small 
because of the low density and high temperature of the halo gas. 
Here the feedback is treated self-consistently 
for both the internal heating of the 
intrahalo gas (within a galaxy's virial radius)
and the external ``pre-heating'' of the surrounding IGM. 
The rarefaction of the pre-heated IGM substantially reduces and even 
reverses the accretion.
The degree of the rarification can be characterized
by the baryonic deficit around a galaxy (e.g., Fig.~\ref{Fig:mf}).

The amount of cooling by the gaseous halo at present can, in principle,
allow for placing constraints on $\eta$. For a too small  $\eta$, 
the gas would not be sufficiently pre-heated to reduce the accretion into 
the halo, and the baryonic mass fraction would be much closer to the
cosmic universal value, as shown in Model VE.
The total cooling rate\footnote{The bolometric luminosity is calculated
based on cooling function with solar abundance \citep{SD93},
and less than 10\% is radiated in 0.3-2.0 keV band.}
of the gas around the reference galaxy
is only a few of $10^{37}{\rm~ergs~s^{-1}}$, the bulk of which arises from
the accumulated reverse-shocked bulge wind material, located outside
of the virial radius (see Fig.~\ref{Fig:lxsb} left panel).
For the model with a lower
$\eta$ value, the cooling would be more efficient and occurs in regions
closer to the bulge.
In general, no over-cooling problem occurs in our model of galaxy evolution
with the proper stellar feedback included.

\subsection{Dynamic state of hot gas}

Probably the most thorough study of hot gas outflows around
ellipticals is by \citet{CiotPel1991}.
\new{They examined the gas dynamics inside a static galactic halo
with a {\it gas free} initial condition that assumes no gas is
present in the galaxy when the type Ia SN feedback begins.
Such a condition may be produced approximately by an intensive
initial SB and possibly AGN feedback in the galaxy central region,
as demonstrated in the present paper.
}
Their 1-D model galaxies evolve through up to
three consecutive evolutionary states: an initial wind,
a transition to subsonic outflow, and a finial inflow phase.
X-ray faint ellipticals are suggested to be in their wind phase,
and brightest ellipticals should mostly be in the inflow regime.
The subsonic outflow is shown to be unstable and short-lived.

Our simulations have shown that the SB feedback is able to produce
a low-density environment near the galactic bulge, which is similar 
to the ``gas-free'' condition assumed in \citet{CiotPel1991}.
But we note that the stellar bulge feedback material
does not leave the galaxy freely.
Much of the feedback energy may be thermalized well inside the
virial radius, forming a hot gaseous halo.
The bulge wind solution applies only within the reverse shock front,
shown by dot-dashed blue lines in Figure~\ref{Fig:mwb4cmp}, for example.
When the reverse shock moves inward, the wind
becomes less and less developed and may eventually turn into
a global subsonic outflow.
We find that this subsonic outflow is stable, 
in contrast to the result of \citet{CiotPel1991}. 
The unstable and short life of the subsonic outflow in
their model is apparently caused by the use of the ``outflow''
outer boundary condition, which is generally not consistent with
the changing properties of the galaxy and its environment. 
The ``outflow'' boundary condition has no effect on the internal flow, 
as long as the outflow is supersonic. But if subsonic, however,
the outflow tends to be suppressed by the artificial force inserted by
the leveled off pressure at the boundary.
The same force would amplify an inflow
and quickly quench the inner bulge wind/outflow.
In our model, no boundary condition needs to be
assumed for the bulge wind/outflow, because it occurs well inside
the spatial range of the simulations. 

The stability of the subsonic outflow state has strong implications for
understanding the large scattering in $L_X/L_B$ ratios of galactic bulges.
The distribution of the hot gas in the state depends 
not only on the feedback in the bulge, but also on the condition
in outer regions of the galactic halo, which is a manifestation of
the large-scale environment and formation history of a galaxy.
This dependence can create a large diversity in the X-ray luminosities
of such galaxies. 
\new{The X-ray luminosity of a galaxy during
the subsonic outflow phase, for example, increases with time because
mass is accumulating in and around the galactic bulge. Therefore, one
may expect a positive correlation between the galaxy age and X-ray
luminosity for galaxies. In addition, the X-ray luminosity of a galaxy
in a higher density environment is expected to be higher. Indeed such
correlations are observationally indicated (e.g., \citealt{OS2001,OP04}).
However, published results on X-ray luminosities from
elliptical galaxies typically suffer from substantial confusion
with stellar contributions (primarily cataclysmic variables and
coronally active binaries; \citealt[and references therein]{Rev2008}).
Careful data analysis similar to that undertaken for the M31 bulge
(e.g., \citealt{Li07b}) and comparison for
the dynamic state of the hot gas are needed to test our explanation for
the large dispersion of the X-ray luminosities.
} 

\subsection{The Missing Stellar Feedback Problem}

For the bulge outflow to be the solution of the missing feedback problem, 
a sufficient mechanical energy output from both the SB and Type Ia SNe is needed.
A weak SB (e.g., a too small $\eta$ value) would not create an environment with
a low enough density for a lasting galactic bulge outflow.
Even with a large SB feedback, a too low SNe Ia rate would lead to the
blowout gas being re-accreted into the halo.
Once an inflow is developed near the bulge, the feedback
from SNe Ia would be consumed locally, producing a luminosity
comparable to the input, which is disfavored by current observations. 

While our 1-D model may provide a reasonably good characterization
of the large-scale impact of the stellar bulge feedback,
the bulge wind/outflow is inefficient in producing diffuse X-ray emission.
In particular, we find that the X-ray luminosity from the 1-D wind/outflow
is considerably lower than what is observed from galactic bulges.
Observationally the reported ``diffuse'' X-ray emission of our Galactic
bulge and the one in M\,31 are about $10^{38}\rm~ergs~s^{-1}$
\citep{Shirey2001,TOKM2004,Li07b}.
For all the models, the X-ray emission from the inner region
is no more than a few of $10^{36}\rm~ergs~s^{-1}$.
Although $\sim 60\%$ of the energy is radiated within half of the
virial radius in Model VE, its X-ray emission near the bulge
region is only a few times $10^{36}\rm ergs~s^{-1}$.
Note that in Model VE about 40\% of the total energy input is used to
remove the stellar mass ejecta out of the gravitational well at $z=0$
and a less fraction is required at early time. It is also worth pointing out
that the diffuse X-ray surface brightness of a bulge wind
is generally steeper than that of stellar mass/light as demonstrated in 
Fig.~\ref{Fig:lxsb} right panel (see also \citealt[fig.~6]{CiotPel1991}).
Only Model VE, in which the gas flow is now in a subsonic outflow
state, has a flatter surface brightness, consistent with observations
of X-ray-faint early-type galaxies (e.g., \citealt{OP04}).

The X-ray faintness of the bulge wind could be
caused by our over-simplified 1-D model.
A careful analysis of the X-ray emission from the M\,31 bulge indicates
the presence of a truly diffuse component with a 0.5-2 keV
luminosity of $\sim 2 \times 10^{38}{\rm~ergs~s^{-1}}$ (Li \& Wang 2007).
However, this component has a substantially lower temperature
($\sim 0.3$\,keV) and a bipolar morphology in and around the bulge.
These facts indicate that much of this soft X-ray component arises from
the interaction of hot gas with cool gas and that the 1-D approximation
is too simplistic, at least in the immediate vicinity of the bulge region, which
may be expected because of the presence of cool gas in the galactic disk.
A detailed study shows that this interaction may indeed be important,
especially for the calculation of X-ray emission, which is sensitive to
presence of density structures. We are carrying out detailed 2-D and 3-D 
simulations of the M31 bulge region and comparisons with corresponding
X-ray observations. The results will be reported in separate papers.

\subsection{Additional Remarks and Caveats}

Stellar feedback is an important process in 
the co-evolution of galactic bulge and host galaxy. 
The two-stage stellar feedback model we have introduced in this paper 
naturally sets up the stellar feedback from a galactic bulge
in the context of the assembly history of the host galaxy/halo.
In the first stage, violent protogalaxy mergers in the fast-assembly
regime trigger a starburst which quickly
releases feedback energy through core-collapsed explosions 
to expel a large fraction of the circum-galactic gas; in the second stage,  
the old stellar population in the bulge provides a long-lasting  
feedback via Type Ia SNe to keep the gaseous halo hot. 
In this model, the SB feedback and 
the gradual feedback are roughly synchronized with the early 
violent merger stage and the later gentle accretion stage of 
the host galaxy, which is a generic picture of galaxy 
formation in the CDM model. A similar co-evolution scenario has 
been proposed for AGN feedback \citep{crot06}. In their 
scenario, a central massive black hole builds up its mass during 
galaxy mergers, and the black hole appears as a ``quasar''; 
when a static hot halo forms in the latter time, the cooling 
halo gas fuels the central massive black hole to make it active 
in a so-called low-energy radio mode.
Although our stellar feedback model and 
the AGN feedback scenario are aimed at two different processes 
in galaxy formation, they share some common features. Firstly, 
the feedback processes happen in two distinct regimes, an early 
energetic but short-term blow-out and a late low-energy level 
but long-lasting heating. Secondly, the feedback and  
the galaxy formation are modeled in a co-evolving fashion. 
However, our stellar feedback model is proposed for the 
MW-sized galaxies, whereas the AGN scenario is
for more massive galaxies, which contribute to
the bright end of the galaxy luminosity function. 

Much of existing work on stellar feedback is focused on whether
a SB could drive a galactic superwind
(e.g., \citealt{MacLow99,Silk03,Fu2004}).
In the model we have examined here, the question whether a SB-driven 
outflow can escape or not is not particularly relevant,
because the galaxy is evolving and the galaxy environment is not gas-free.
More relevant is the cooling time scale of the SB-heated material.
This cooling time depends on the detailed physical 
processes involved in the interplay between massive stars and the ISM, 
which our simple model does not take into account. 
Depending largely on the assumed value of $\eta$, the SB-heated 
material may have already fallen back to the center or is still 
hanging in regions outside the galactic virial radius.
Our choice of the initial condition of the SB material
(the distribution scale) guarantees that the cooling time-scale 
is long enough for the Type Ia SN-driven wind to develop an outflow.
It is the rarification of the halo gas by the SB feedback that helps
the Type Ia SN-driven outflow to gently blow out the accreted material
and replenish the ISM with mass from evolving stars.

We mainly focus on a simultaneous solution to solve the
``over-cooling'' and ``missing energy'' problems in 
intermediate-mass ($\sim 10^{12}\rm\,M_\odot$) galaxies with
a substantial stellar bulge component.
The combination of the SB and gradual stellar feedbacks is demonstrated to
be a viable way to solve the problems.
In the following we discuss briefly the limitations of our model.

How galactic stellar bulges really form remains a question of debate.
It is likely that there are multiple formation paths.
A popular paradigm is the formation through galaxy merges (instead
of a monolithic collapse), which
typically happen at high redshifts. In fact, our halo growth model is based on
the average mass accretion history from $N$-body simulations
(e.g., \citealt{Wechsler2002,Zhao03a,Zhao03b}). 
After the last major merger, a bulge evolves nearly passively, 
except for some minor mergers. 
If the bulk of bulge stars are formed in a SB during the last major
merge, our model still applies. However, if the bulge stars forms primarily
from pre-existing stars by mergers (the so-called dry mergers),
a direct application of our model would not be appropriate.

We have considered galaxies  in ``isolation''.  
This approximation may be reasonable for field galaxies, especially
at high redshifts. But as shown in our simulations, 
the stellar feedback may have significant effects
on scales considerably larger than the virial radius of
individual galaxies. Such impact may be sufficient to influence 
the IGM environments of member galaxies in groups. For example, much
of the gas in the Local Group, which contains two MW-like galaxies with a 
substantial stellar bulge component, may have been heated to a temperature
of $\sim 10^7$ K (e.g., Fig.~\ref{Fig:mwb4a}). 
Such hot gas with too low a density to be easily detected
is consistent with the existing observations. 
But the stellar energy feedback in groups containing
galaxies with more massive 
halos (e.g., giant ellipticals) may have relatively minor effects,
compared to the gravitational heating. 

In our numerical model, mass accretion is assumed to be smooth.
In reality, however, the accretion is more complicated. 
$N$-body simulations of cosmic structure formation have shown 
that dark matter halos acquire their masses mainly through lumpy 
mergers rather than smooth accretion. The SPH galaxy formation 
simulations have further indicated that the gas 
accretion comes from both a smooth ``hot'' mode and a clumpy ``cold'' 
mode \citep{Keres05}. As discussed in \citet{voit03}, 
lumpy accretions may give rise to higher 
entropy in the IGM than smoothed accretion,
and hence may result in less cooling. In addition, since 
denser clumps are difficult to be shock-heated (e.g. \citealt{voit03, pool06}),
the gas accretion in different modes may lead to a multi-phase
medium, which  may alter the assembly  
history of the central galaxy, and hence changes the evolution 
of the stellar feedback and the thermal properties of the halo 
gas \citep{Mo96, Maller04, kauf06}. While the details of such
a multi-phase medium are still poorly understood,
the clumpy accretion alone cannot solve the
``overcooling'' and ``missing energy'' problems,
unless additional pre-heating is injected into the gas before
it is  accreted (e.g., \citealt{Lu07}).
The expectation is that the multi-phase medium may increase the 
gas accretion rate onto the central galaxy, hence boosting the 
predicted X-ray luminosity. The cold accretion may also be responsible 
to some of the observed HVCs around our Galaxy 
(e.g., \citealt{sem03}), in addition to those 
that might be produced by the cooled feedback material
discussed in \S3.1.4.
\new{The observed HVCs show a range of metallicity with the mean of
$\sim 0.1$ solar, as is kind of expected from the initial SB chemical
enrichment assumed here and is often used as a canonical value, 
although the number of good measurements is still very limited. While 
this consistency is encouraging, more work is needed to understand the 
HVC production in our scenario, probably based on careful 3-D 
simulations, and on a systematic comparison between simulations
and observations in terms of mass and velocity distributions, for example.
}

For spiral galaxies, additional cool gas may be channeled into the
bulge regions via gravitational perturbations because of the presence
of a stellar bar. The heating of the gas (e.g., by thermal conduction)
may lead to mass-loading of the hot gas,
increasing the X-ray luminosity and reducing the temperature of the gas.
We will study such effects in a separate paper.

\color{black}

\section{Summary}

We have examined the potential impacts of the galactic stellar bulge feedback 
on the hot gaseous structure and evolution around intermediate-mass galaxies
with substantial bulge component.
These are typically early-type spirals such as the MW,  M\,31
and ellipticals with low $L_X/L_B$ ratios, indicating
ongoing mass and energy outflows. The feedback in such a galaxy 
is approximately divided
into two phases: 1) a SB-induced blastwave from the formation of the
bulge and 2) a gradual energy and mass injection from evolved bulge stars.
Our 1-D model deals with the feedback
in the galaxy evolution context, approximately accounting for the
dark matter accretion history. We have carried out various simulations
with this model to demonstrate the impacts and their dependences on
key assumptions and parameters used. Our major results and conclusions 
are as follows:

\begin{itemize}
\item The blastwave initiated during the SB and maintained by subsequent 
Type Ia SNe in a galactic bulge may heat the IGM beyond the virial radius, 
which may stop further gas accretion. The heated gas expands and produces 
a baryon deficit around the galaxy. 

\item The high temperature and low density of the remaining 
gas in the galactic halo produces little X-ray emission, consistent with
existing observations. Thus the inclusion of both the SB and Type Ia SN
feedbacks may provide a solution to the theoretical over-cooling problem.

\item With the low-density circum-galactic gaseous environment,
the gradual stellar feedback, powered by Type Ia SNe alone, can form a
galactic bulge wind, especially when the bulge is young. Much of the
tenuous hot halo is occupied by the thermalized bulge wind material.
The bulge wind region shrinks and likely turns into a subsonic outflow
as the SN rate decreases. This wind/outflow naturally explains the 
{\sl missing stellar feedback} problem of galactic bulges. 

\item We find that a subsonic outflow from a galactic bulge can be
a long-lasting state, the properties of which depend sensitively on
the environment and formation history of the galaxy. This dependency
may account for the large dispersion in the X-ray 
luminosities of galaxies with similar $L_B$.

\item The condensation of the gas injected during the SB may provide
an explanation of HVCs observed with moderate metal enrichment.

\end{itemize}

\section*{Acknowledgments}
We thank the referee for useful comments that led to
improved presentation of this paper.
The software used in this work was in part developed by the DOE-supported
ASC / Alliance Center for Astrophysical Thermonuclear Flashes at
the University of Chicago. This project is supported by
NASA through grant SAO TM7-8005X and NNG07AH28G.
HJM  would like to acknowledge the support
of NSF AST-0607535, NASA  AISR-126270 and NSF IIS-0611948.


\begin{figure}[htbp]
\begin{center}
\plotone{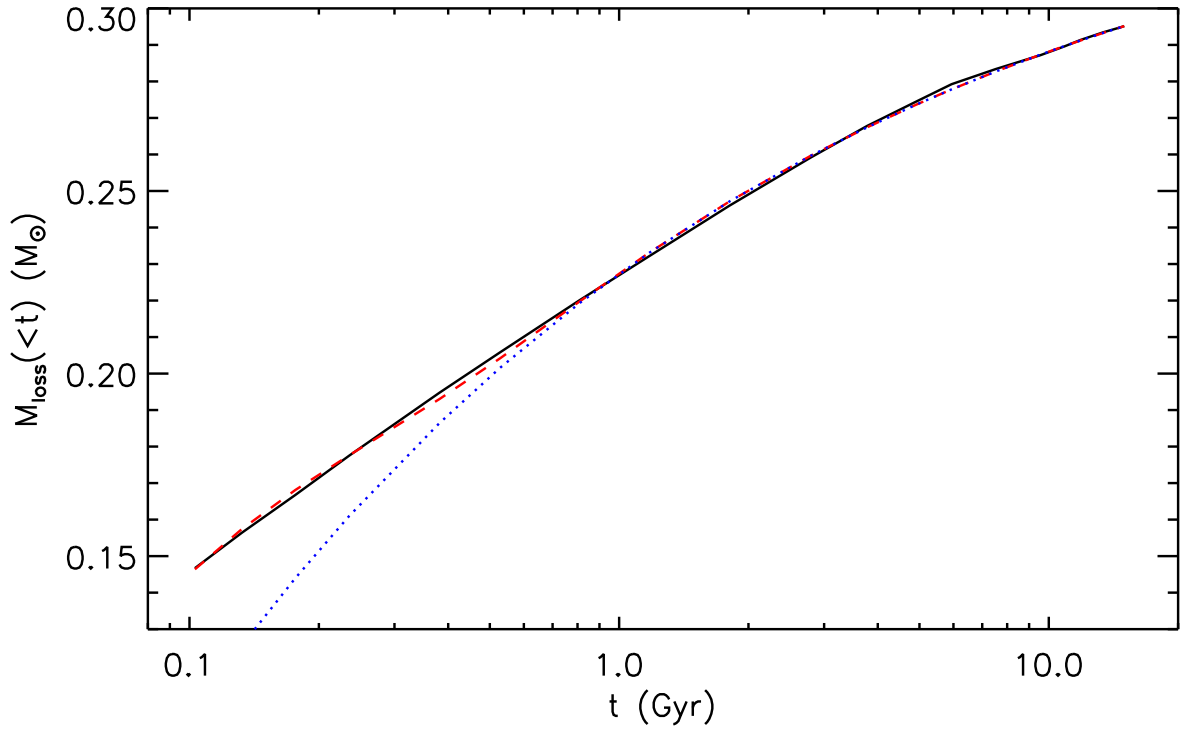}
\caption{\label{Fig:fitmlr}
  The cumulative stellar mass loss of a stellar population
  (normalized to one $\msun$ with the Salpeter IMF)
  as a function of its age.
  The solid black line denotes the data taken from Maraston (2005)
  Figure 22.; the dashed red line is the fitting; the dotted blue line
  is the component with the same power index as used in Ciotti (1991).
}
\end{center}
\end{figure}

\epsscale{0.9} 
\begin{figure}[htbp]
\begin{center}
\plotone{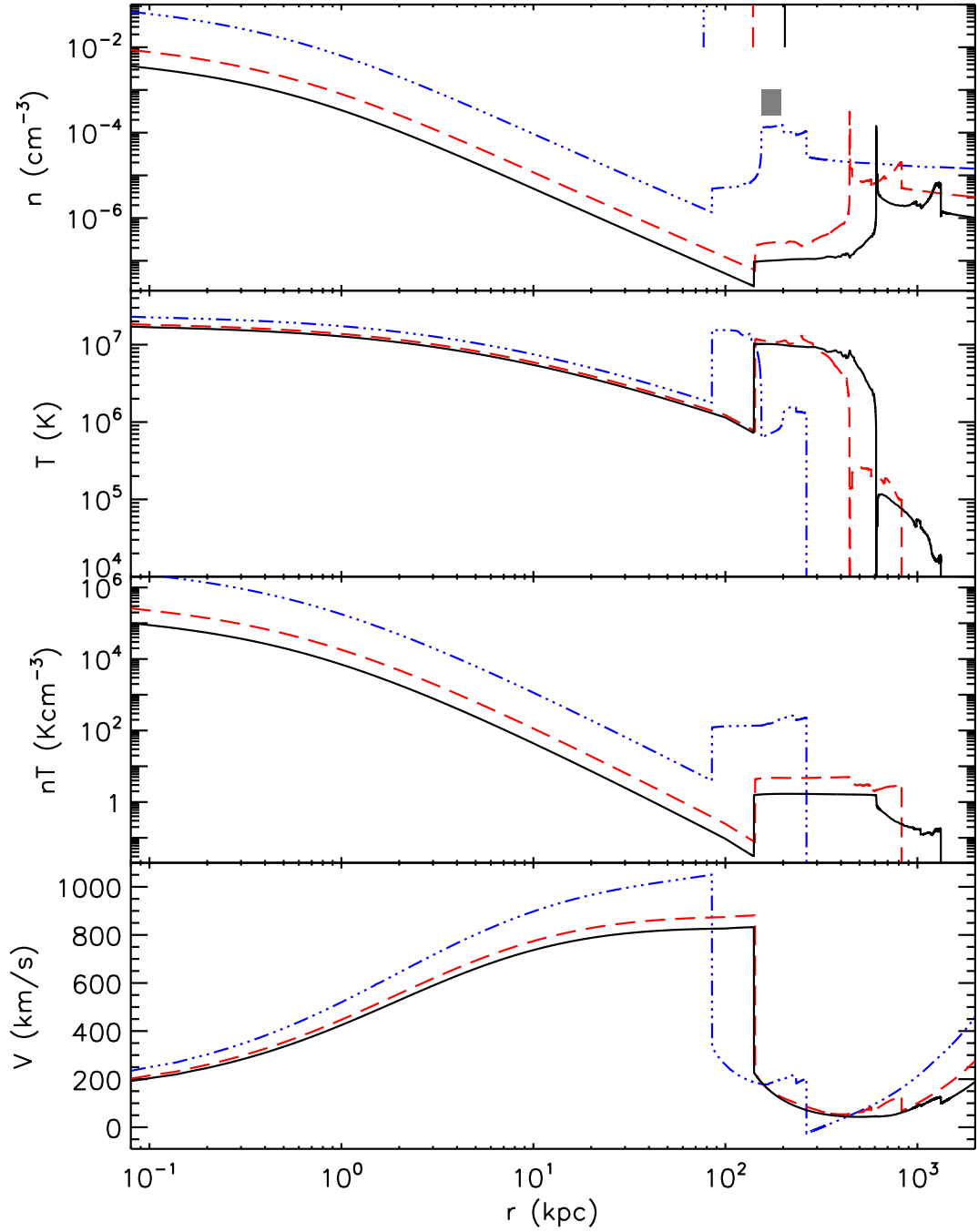}
\caption{\label{Fig:mwb4a}
Radial profiles of density, temperature, pressure, and velocity of the gas
around the simulated reference galaxy at three representative redshifts: 
$z$=1.4 (dot-dashed blue lines), 0.5 (long-dashed red), 
and 0.0 (solid black). 
The filled gray region in the top panel marks the radial range of 
the SB-enriched material at $z$=1.4. In the lower $z$ profiles, this range has 
collapsed and is seen as the narrow spikes.
\new{Vertical lines in the uppermost panel indicate where the virial radius lies at
a given color-coded redshift. Those lines are drawn in other plots with the
same meaning.}
}
\end{center}
\end{figure}
\epsscale{1.0}

\begin{figure}[htbp]
\begin{center}
\plotone{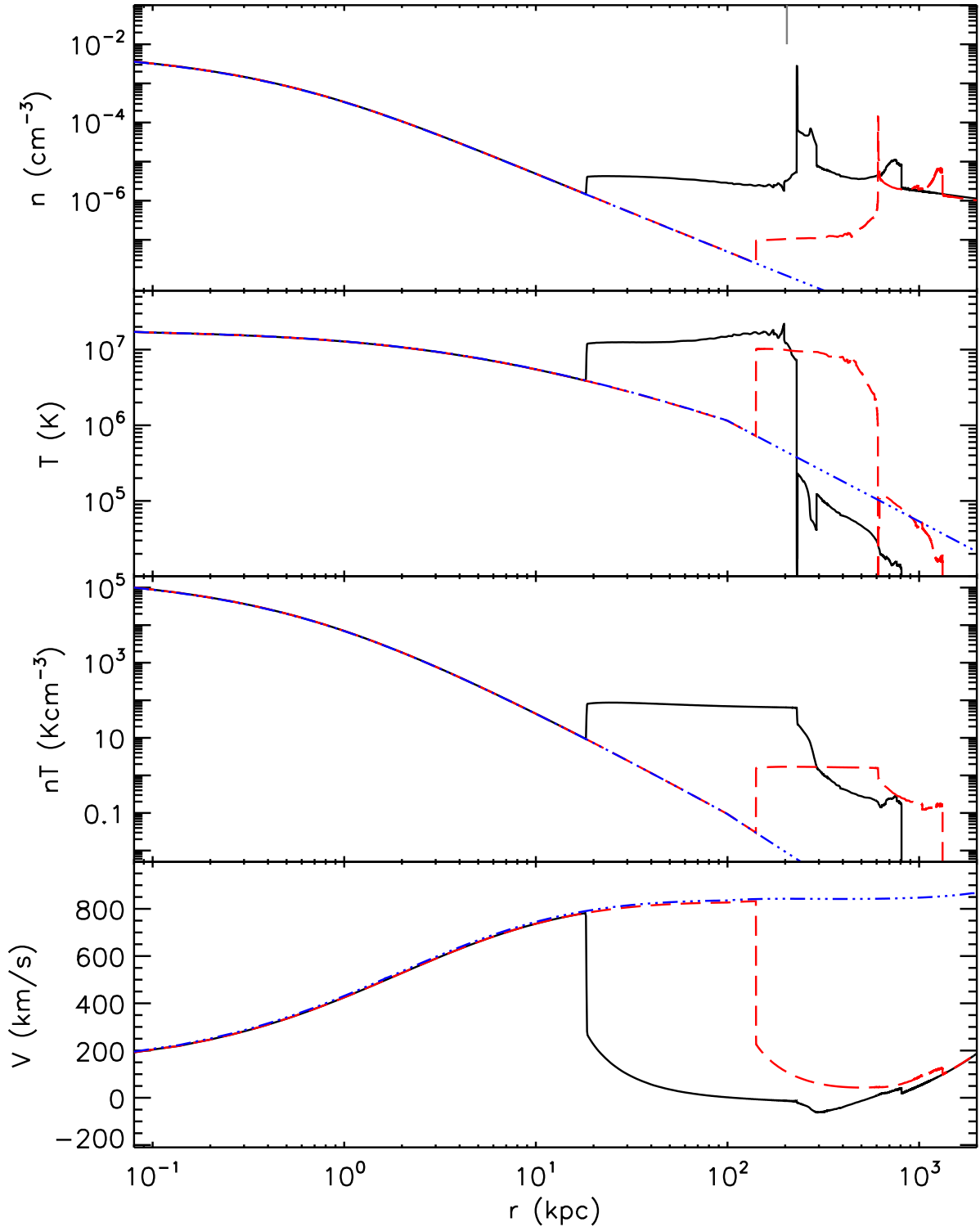}
\caption{\label{Fig:mwb4cmp}
  Radial profiles at $z$=0 for three models: the reference model
  (long-dash red lines); Model VA (solid black lines);
  and a model started with the {\sl gas free} initial condition
  (three-dot-dashed blue lines).
  \new{The gas-free initial condition corresponds to that
  of Ciotti et al. (1991), and is discussed in section 4.2.
}
  The noticeable wiggles in the profiles of temperature and density
  are generated by the unstable cooling when the dense shell of
  SB material forms.
}
\end{center}
\end{figure}

\begin{figure}[htbp]
\begin{center}
\plotone{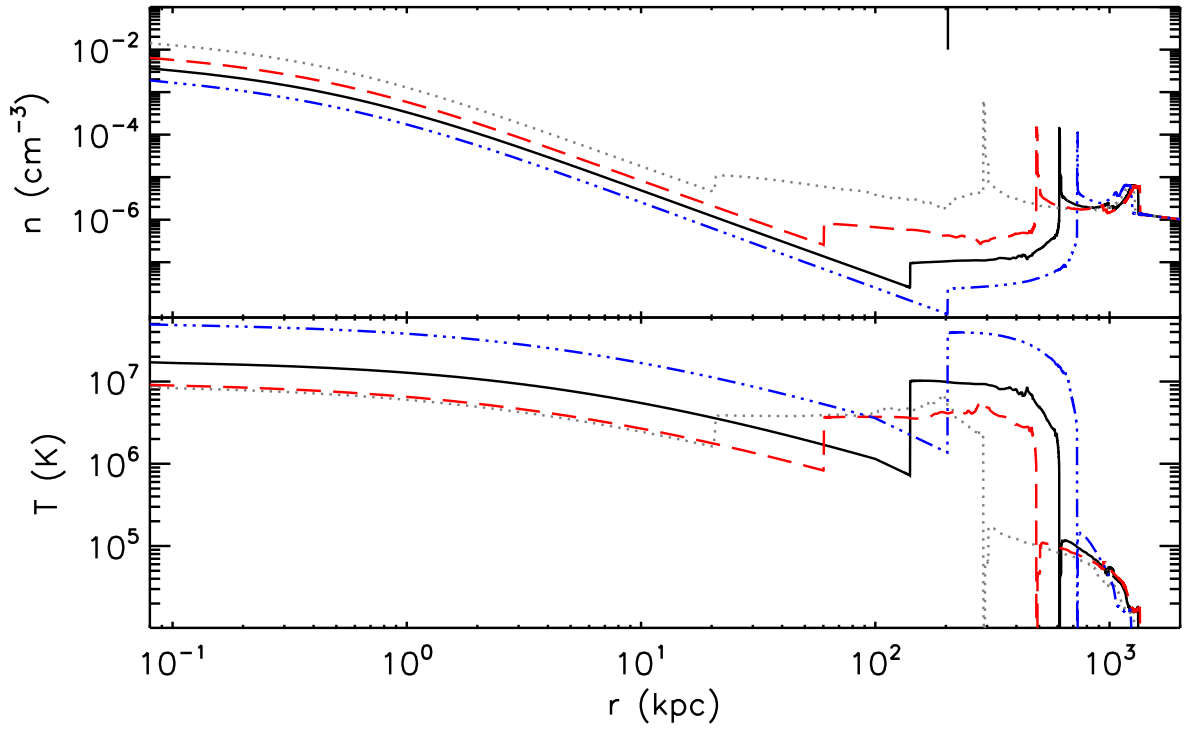}
\caption{\label{Fig:cmpp}
  Comparison of the radial density and temperature profiles inferred
  from the models with various gradual feedback parameters at $z=0$:
  the reference model (solid black lines),
  model VB (dotted grey lines),
  model VC (dashed red lines),
  model VD (three-dot-dashed blue lines).
}
\end{center}
\end{figure}

\begin{figure}[htbp]
\begin{center}
\plotone{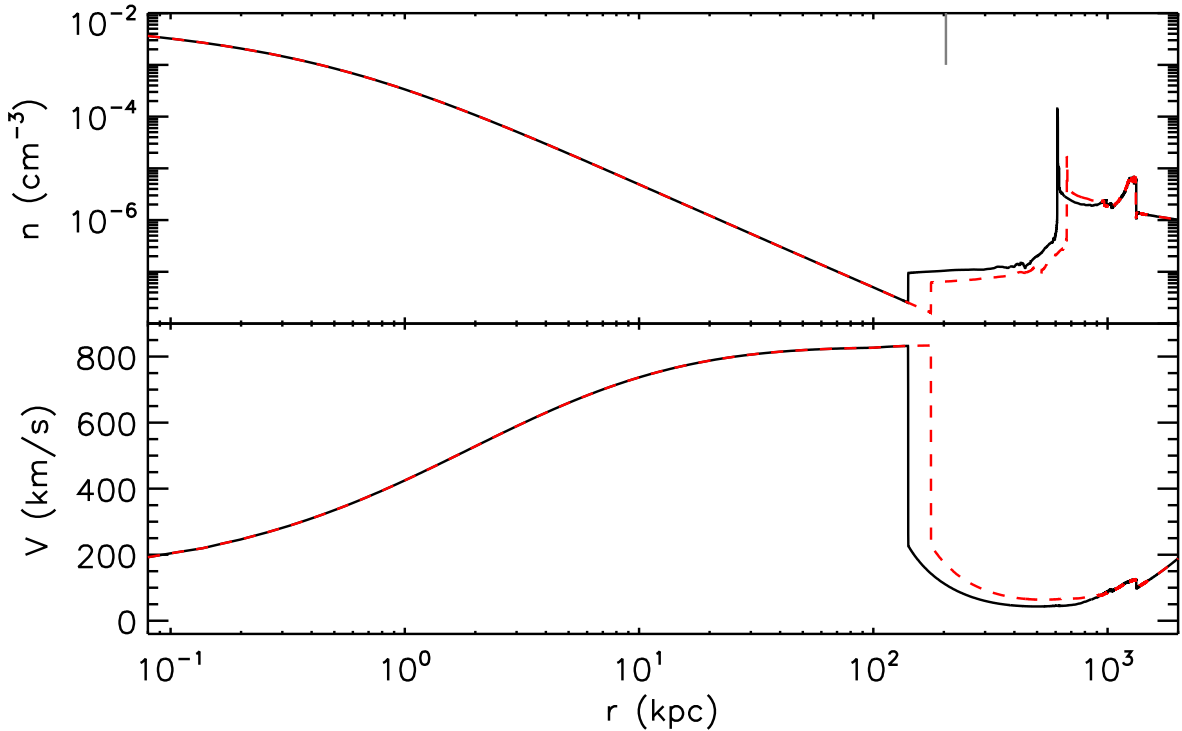}
\caption{\label{Fig:cmpdo}
  Demonstration of the ``dropout'' effect on the reference model
  at $z$=0: the reference model (solid black lines),
  the reference model with the ``dropout'' included (dashed red lines).
}
\end{center}
\end{figure}

\begin{figure}[htbp]
\begin{center}
\plotone{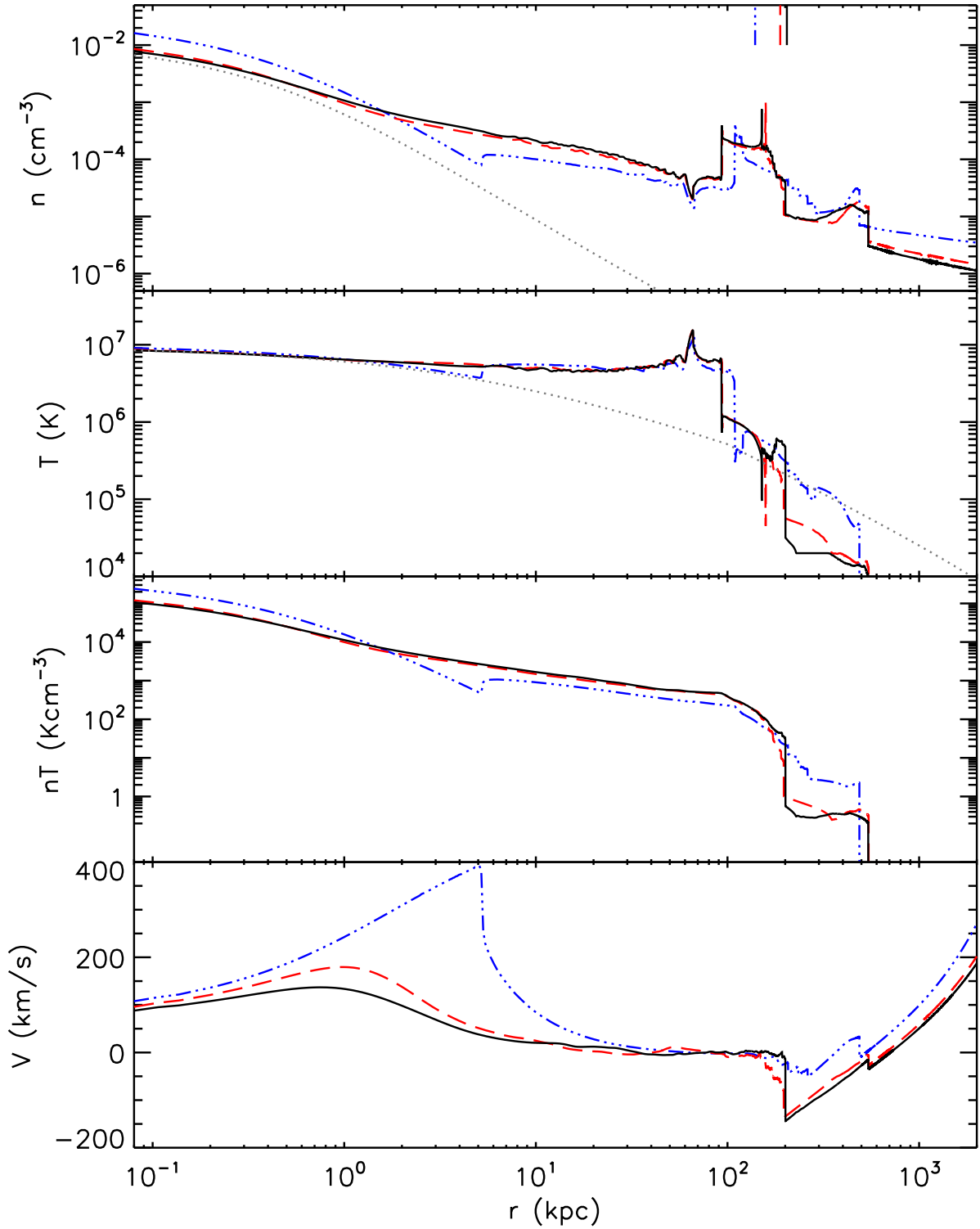}
\caption{\label{Fig:stado}
  Radial profiles of Model VE (with the ``dropout'' enabled)
  at three redshifts: $z$=0.5 (three-dot-dashed blue line),
  $z$=0.1 (dashed red lines), and $z$=0.0 (solid black lines).
  The dotted grey lines in the top two panels 
  denote the corresponding profiles if model VE starts
  from a {\sl gas free} condition.
}
\end{center}
\end{figure}

\begin{figure}[htbp]
\begin{center}
\centerline{
\psfig{file=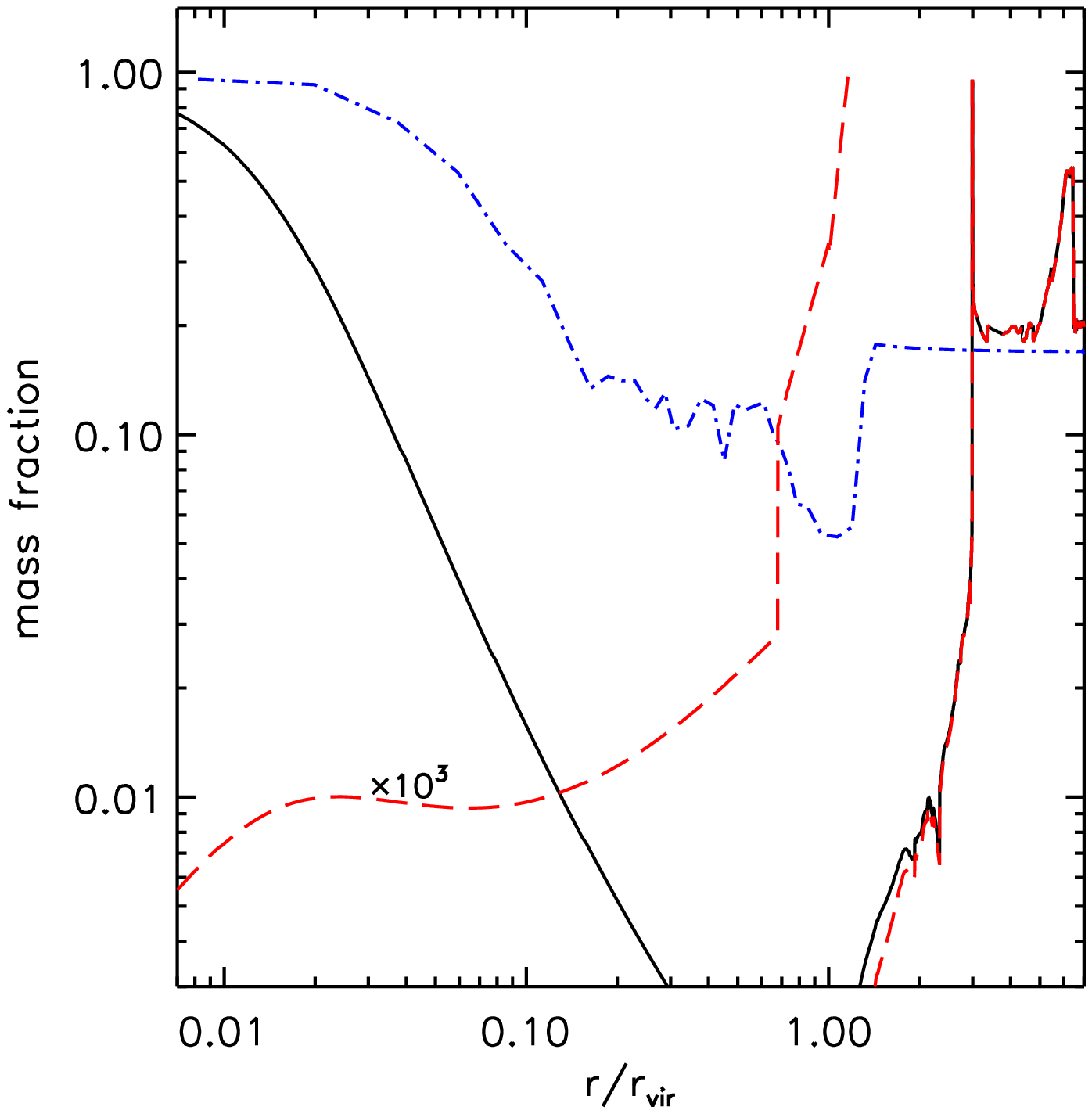,width=0.3\textwidth}
\hspace{0.1in}
\psfig{file=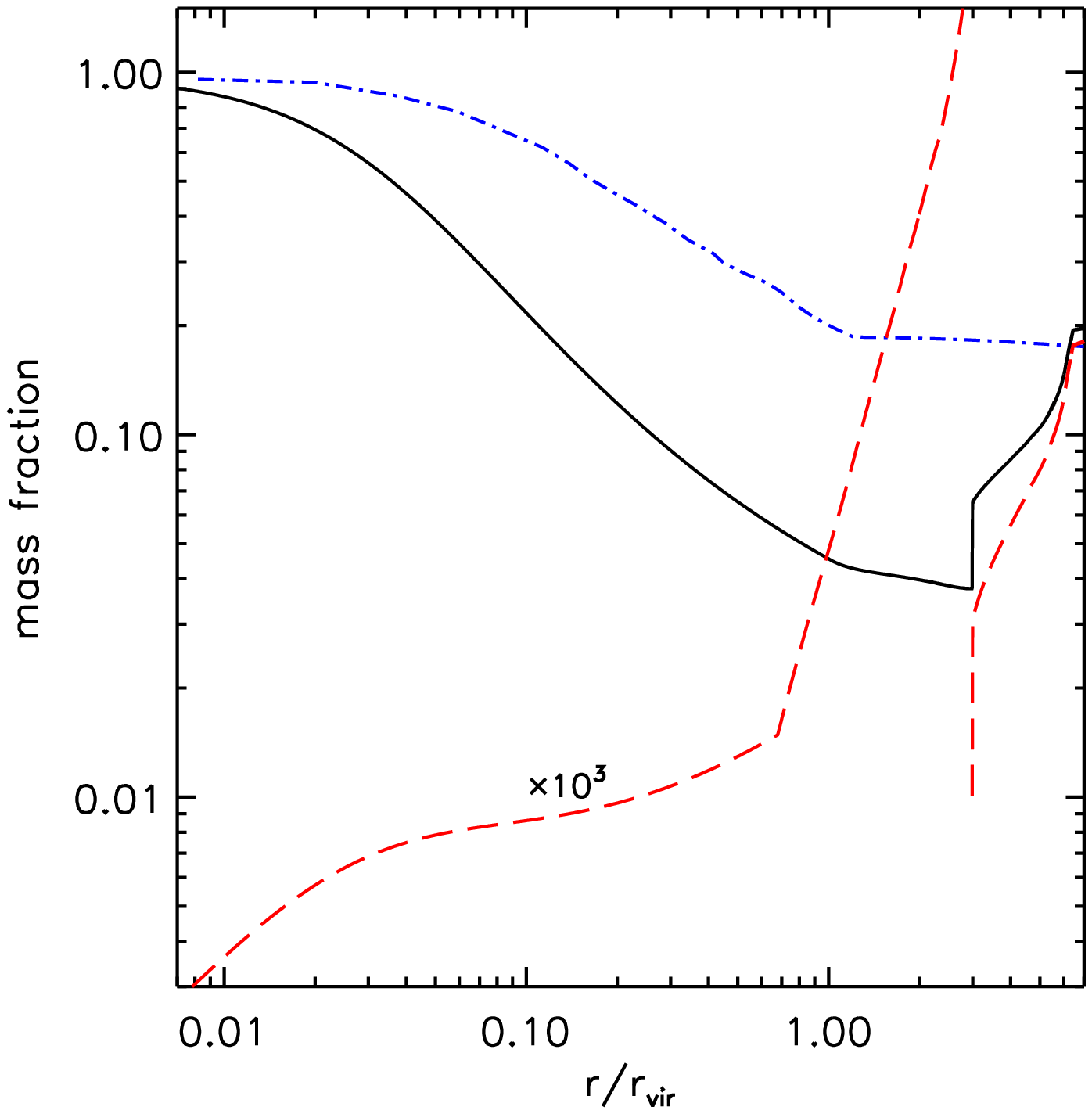,width=0.3\textwidth}
\hspace{0.1in}
\psfig{file=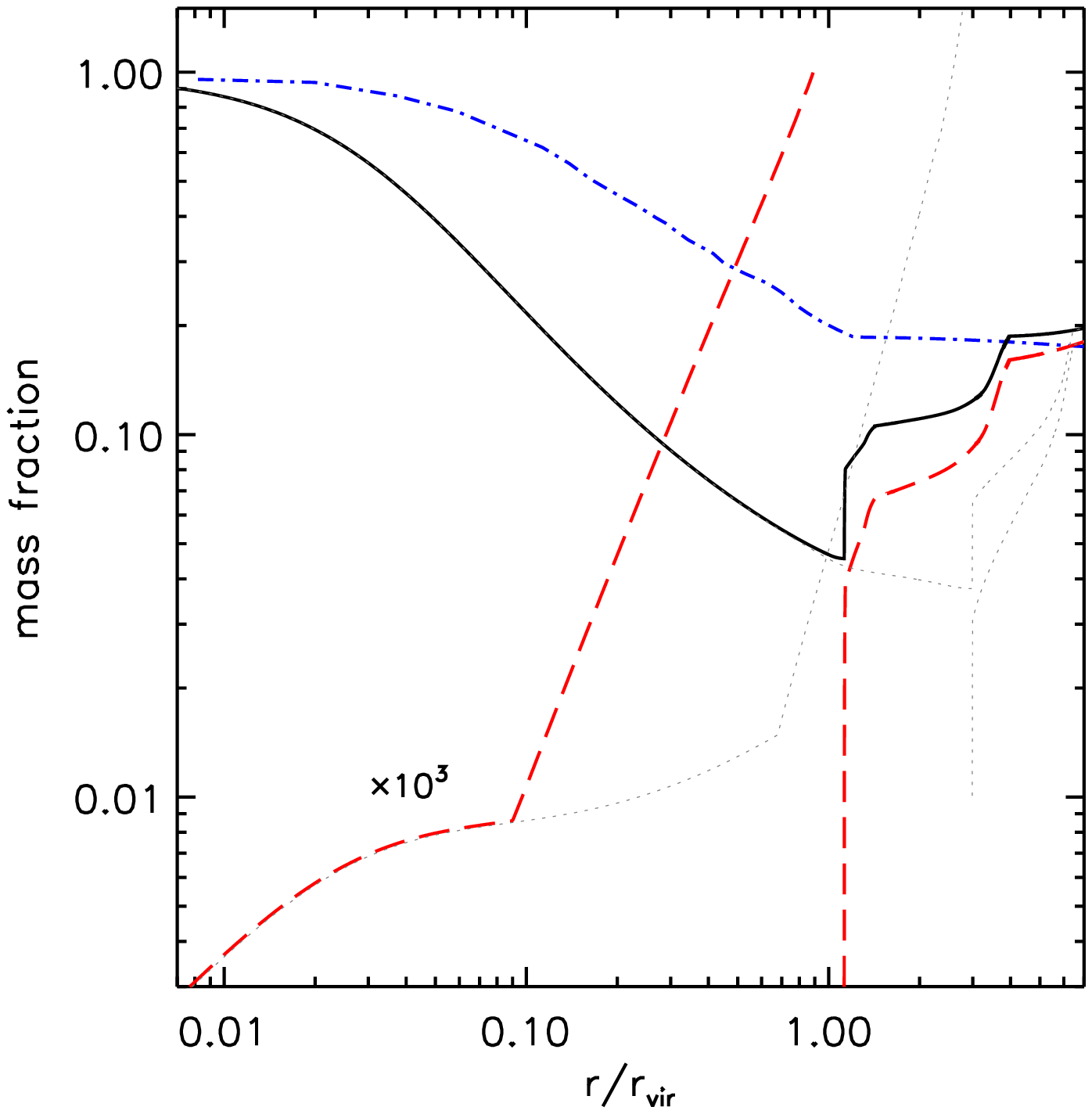,width=0.3\textwidth}
}
\caption{\label{Fig:mf}
  The local baryonic fraction as a function of the dimensionless
  galactocentric radius in the unit of virial radius for 
  the reference model ({\sl left panel});
  the accumulated baryonic fraction as a function of the radius
  for the reference model ({\sl middle panel}) and Model VA (Table 2,
  {\sl right panel}).
  The dot-dashed blue lines are for the total baryonic
  mass fraction at the starburst time $t_{sb}$.
  The solid black lines are for the total baryonic  mass fraction at $z$=0.
  Dashed red lines denote the contribution of the hot gas,
  with the inner part increased by a factor of $10^3$
  for ease visualization. The baryonic fractions reach to the universal value 
  $f_{b}$=0.17 near the forward shock front located at a few times
  of the virial radius.
 \new{The dotted gray lines in the right panel denote the
   corresponding results of the reference model for directly
   comparison with Model VA.}
  }
\end{center}
\end{figure}

\begin{figure}[htbp]
\begin{center}
\plottwo{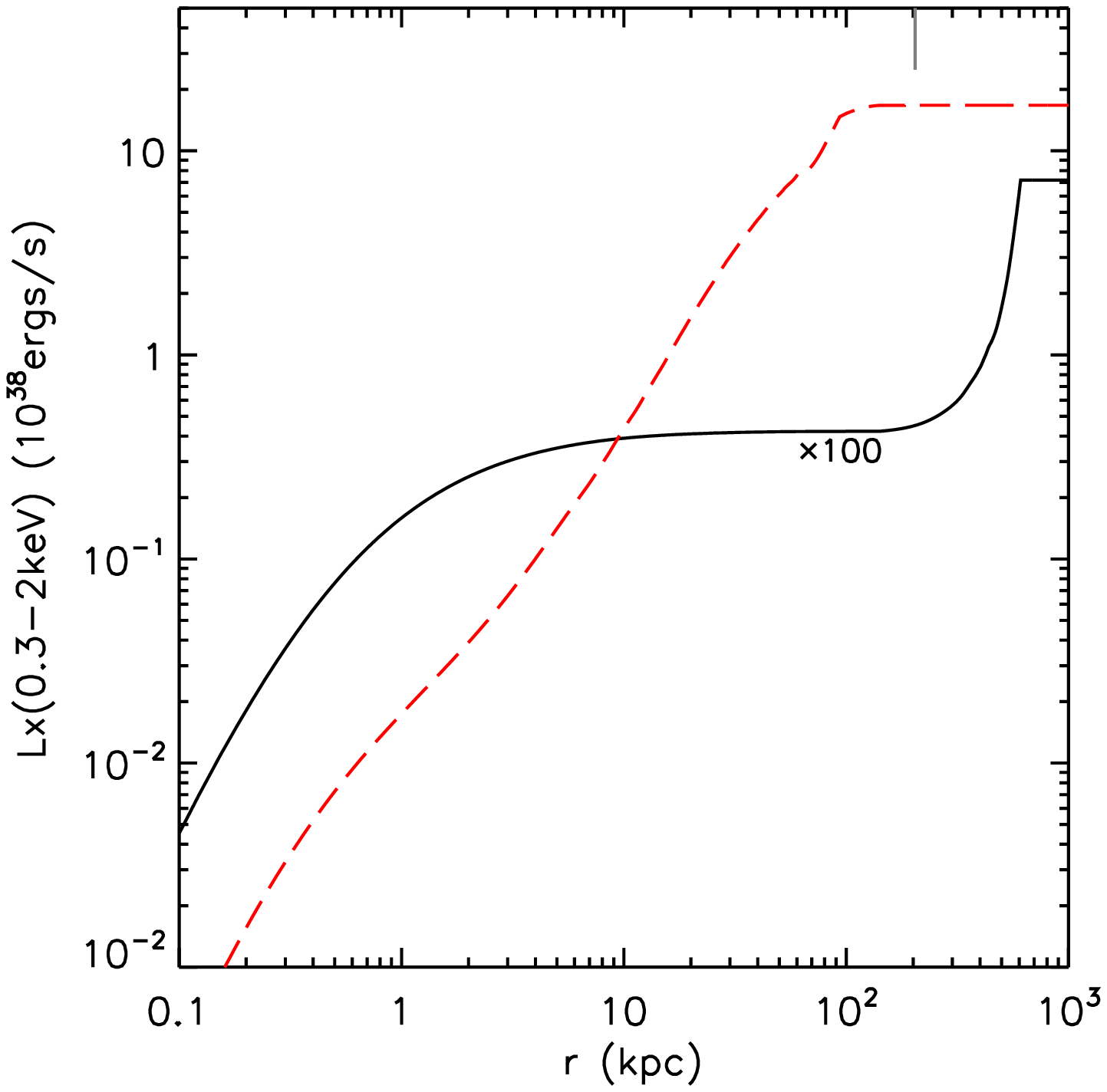}{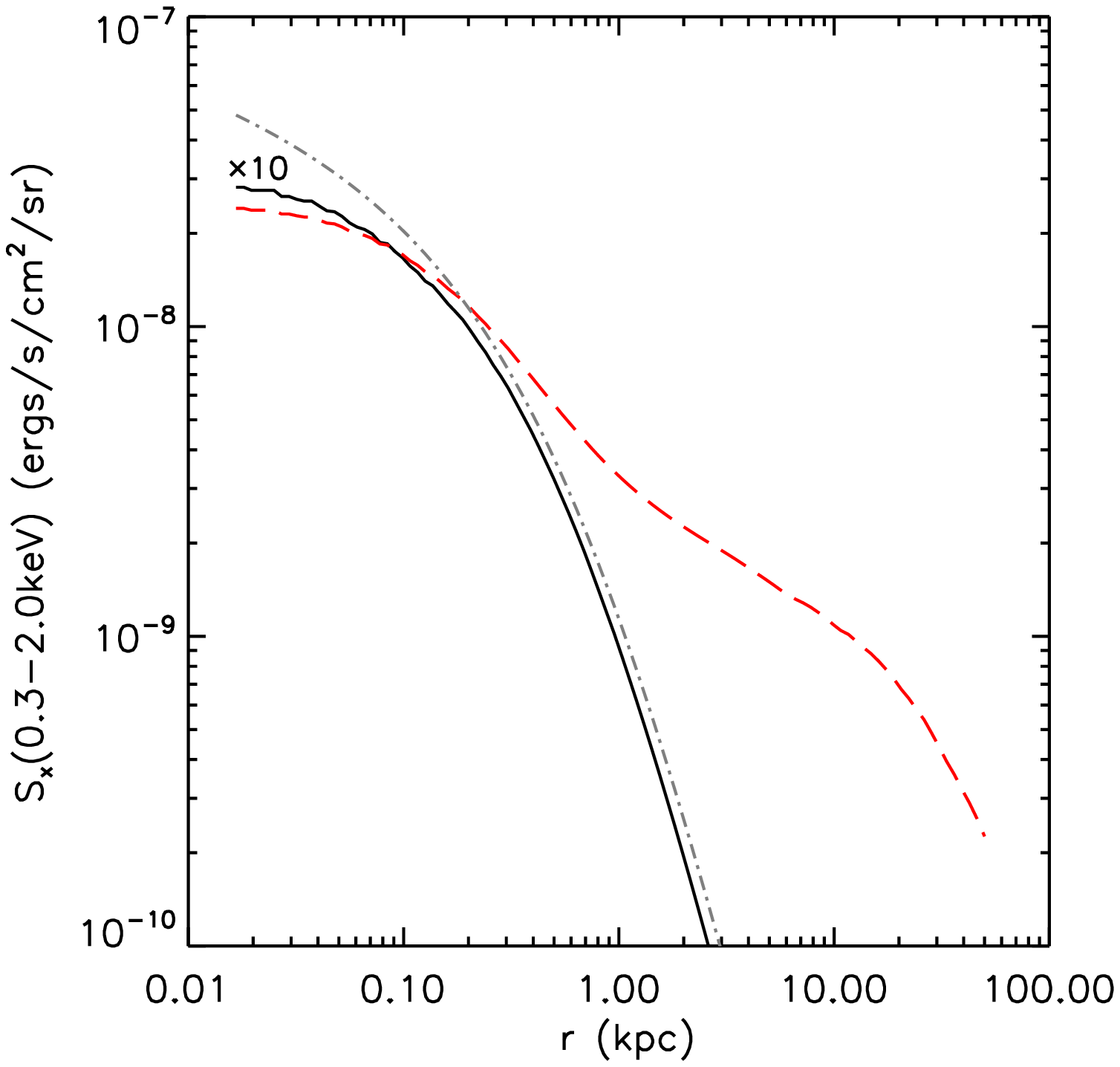}
\caption{\label{Fig:lxsb}
  {\sl Left Panel:} Cumulative X-ray luminosities as functions of galactocentric
  radius in the 0.3-2.0 keV band for
  the reference model (solid black line) and the model VE
  (solid black line).
  {\sl Right panel:} The corresponding surface brightness of the X-ray emission,
the dot-dashed gray line denotes
  the stellar surface brightness with an arbitrary normalization.
  The values of the reference model are multiplied by the marked factor
  for easy visualization.
}
\end{center}
\end{figure}


\begin{thebibliography}{}
\setlength{\itemsep}{-0.155truecm}

\bibitem[Arthur(1999)]{Ar99}
Arthur N, Cox, 1999, Allen's Astrophysical Quantities, fourth edition

\bibitem[Aubourg et~al.(2007)]{Aub07}
Aubourg, E., Tojeiro, R., Jimenez, R., et~al. 2007, astroph/0707.1328v2

\bibitem[Birnboim \& Dekel (2003)]{Bir03}
Birnboim, Y., \& Dekel, A., 2003, MNRAS, 345, 349

\bibitem[Blumenthal et~al.(1986)]{blum86} 
Blumenthal G. R., Faber S. M., Flores R., Primack J. R., 1986, ApJ, 301, 27

\bibitem[Bullock et~al.(2001)]{bull01}
Bullock J. S., Kolatt T. S., Sigad Y., Somerville R. S., Kravtsov A. V., Klypin A. A., Primack J. R., Dekel A., 2001, MNRAS, 321, 559

\bibitem[Bond \etal(1991)]{bond91} 
Bond J. R., Cole S., Efstathiou G., Kaiser N., 1991, ApJ, 379, 440

\bibitem[Brighenti \& Mathews(1999)]{Bri99}
Brighenti F., Mathews W. G., 1999, ApJ, 512, 65

\bibitem[Bryan \&  Norman (1998)]{Bryan98}
Bryan G. L., Norman M. L., 1998, ApJ, 495, 80

\bibitem[Cappellaro et~al(1999)]{Cap99}
Cappellaro E., Evans R., Turatto M., 1999, A\&A, 351, 459

\bibitem[Ciotti et~al.(1991)]{CiotPel1991}
Ciotti L., D'Ercole A., Pellegrini S., Renzini A., 1991, ApJ, 376, 380

\bibitem[Choi et~al.(2006)]{choi06}
Choi J., Lu Y., Mo H. J., Weinberg M. D., 2006, MNRAS, 372, 1869

\bibitem[Croton \etal(2006)]{crot06} 
Croton D. J., et al., 2006, MNRAS, 365, 11

\bibitem[Dav\'e \& Oppenheimer(2007)]{DO07}	
Dav\'e, R., Oppenheimer, B. D., 2007, MNRAS, 374, 427

\bibitem[David et~al.(1991)]{David1991}
David L. P., Forman W., Jones C., 1990, ApJ, 359, 29

\bibitem[David et~al.(2006)]{David2006}
David L. P., Jones C., Forman W., Vargas I. M., Nulsen P., 2006, ApJ, 653, 207




\bibitem[Fryxell et~al(2000)]{Fry00}
Fryxell B., et al., 2000, ApJS, 131, 273

\bibitem[Fujita et~al(2004)]{Fu2004}
Fujita, A., Mac Low, M. M., Ferrara, A., \& Meiksin, A. 2004, ApJ, 613, 159

\bibitem[Fukugita \& Peebles (2006)]{fuk06}
Fukugita M., \& Peebles P. J. E., 2006, ApJ, 639, 590

\bibitem[Gnedin(2000)]{gned00} 
Gnedin N. Y., 2000, ApJ, 542, 535

\bibitem[Hoeft \etal(2006)]{hoef06} 
Hoeft M., Yepes G., Gottl\"ober S., Springel V., 2006, MNRAS, 371, 401

\bibitem[Greggio (2005)]{Greg05}
Greggio L., 2005, A\&A, 441, 1055

\bibitem[Hernquist(1990)]{Her90}
Hernquist L., 1990, ApJ, 356, 359

\bibitem[Jeltema \etal(2008)]{Jeltema08}
Jeltema T. E., Binder B., Mulchaey J. S., 2008, ApJ, 679, 1162

\bibitem[Jing \& Suto(2000)]{jing00} 
Jing Y. P., Suto Y., 2000, ApJ, 529, L69

\bibitem[Kaufmann et~al(2006)]{kauf06}
Kaufmann T., Mayer L., Wadsley J., Stadel J., Moore B., 2006, MNRAS, 370, 1612

\bibitem[Kere\v{s} et~al. (2005)]{Keres05}
Kere\v{s} D., Katz N., Weinberg D. H., Dave R., 2005, MNRAS, 363, 2

\bibitem[Kroupa(2001)]{Kroupa01}
Kroupa P., 2001, MNRAS, 322, 231

\bibitem[Li et~al.(2006)]{Li06}
Li Z., Wang Q. D., Irwin J. A., Chaves T., 2006, MNRAS, 371, 147

\bibitem[Li et~al.(2007)]{Li07a}
Li Z., Wang Q. D., Hameed S., 2007, MNRAS, 376, 960

\bibitem[Li \& Wang (2007)]{Li07b}
Li Z., Wang Q. D., 2007, ApJ, 668, L39

\bibitem[Lowenstein \& Mathews(1987)]{LM1987}
Loewenstein M., Mathews W. G., 1987, ApJ, 319, 614

\bibitem[Lu et~al.(2006)]{Lu06}
Lu Y., Mo H. J., Katz N., Weinberg M. D., 2006, MNRAS, 368, 1931

\bibitem[Lu \& Mo (2007)]{Lu07}
Lu Y., Mo H. J., 2007, MNRAS, 377, 617

\bibitem[MacLow \& Ferrara (1999)]{MacLow99}
Mac Low  Mordecai-Mark, \& Ferrara  Andrea, 1999, apj, 513, 142

\bibitem[Maller \& Bullock (2004)]{Maller04}
Maller A. H., Bullock J. S., 2004, MNRAS, 355, 694

\bibitem[Mannucci \etal(2006)]{Man06}
Mannucci F., Dell V. M., Panagia N., 2006, MNRAS, 370, 773

\bibitem[Maraston (1998)]{Mar98}
Maraston, C. 1998, MNRAS, 300, 872

\bibitem[Maraston(2005)]{Mar05}
Maraston C., 2005, MNRAS, 362, 799

\bibitem[Mathew \& Baker(1971)]{MB1971}
Mathews W. G., Baker J. C., 1971, ApJ, 170, 241
 
\bibitem[McCarthy \etal(2008)]{McCarthy08}
McCarthy I. G., Frenk C. S., Font A. S., Lacey C. G., Bower R. G.,
Mitchell N. L., Balogh M. L., Theuns T., 2008, MNRAS, 383, 593


\bibitem[McGaugh (2007)]{Mc07}
McGaugh S. S., 2007, IAUS, 244, 136

\bibitem[Miller \& Scalo (1997)]{MS79}
Miller G. E., Scalo J. M., 1979, ApJS, 41, 513

\bibitem[Mo \& Mao (2002)]{Mo02}
Mo H. J., Mao S., 2002, MNRAS, 333, 768

\bibitem[Mo \& Mao (2004)]{Mo04}
Mo H. J., Mao S., 2004, MNRAS, 353, 829

\bibitem[Mo \& Miralda-Escude(1996)]{Mo96}
Mo H. J., Miralda-Escude J., 1996, ApJ, 469, 589

\bibitem[Mo \etal (2005)]{Mo05}
Mo H. J., Yang X., van den Bosch, F. C., Katz N., 2005, MNRAS, 363, 1155

\bibitem[Moore \etal(1998)]{moor98} 
Moore B., Governato F., Quinn T., Stadel J., Lake G., 1998, ApJ, 499, L5

\bibitem[Navarro, Frenk \& White(1996)]{nfw96} 
Navarro J. F., Frenk C. S., White S. D. M., 1996, ApJ, 462, 563

\bibitem[O'Sullivan et~al.(2001)]{OS2001}
O'Sullivan Ewan, Forbes Duncan A., Ponman Trevor J., 2001, MNRAS, 324, 420

\bibitem[O'Sullivan \& Ponman(2004)]{OP04}
O'Sullivan E., Ponman T. J., 2004, MNRAS, 349, 535


\bibitem[Pellegrini \& Ciotti(1998)]{Pel98}
Pellegrini S., Ciotti L., 1998, A\&A, 333, 433

\bibitem[Poole \etal(2006)]{pool06} 
Poole G. B., Fardal M. A., Babul A., McCarthy I. G., Quinn T., Wadsley J., 2006, MNRAS, 373, 881

\bibitem[Press \& Schechter(1974)]{pres74} 
Press W. H., Schechter P., 1974, ApJ, 187, 425

\bibitem[Quinn \etal(1996)]{quin96}
Quinn T., Katz N., Efstathiou G., 1996, MNRAS, 278, L49

\bibitem[Revnivtsev \etal(2006)]{Rev06}
Revnivtsev M., Sazonov S., Gilfanov M., Churazov E., Sunyaev R., 2006, A\&A, 452, 169

\bibitem[Revnivtsev et~al.(2008)]{Rev2008}
Revnivtsev M., Molkov S., Sazonov S., 2008, A\&A, 483, 425

\bibitem[Schaerer(2003)]{Schaerer03}
Schaerer D., 2003, A\&A, 397, 527

\bibitem[Sembach \etal (2003)]{sem03}
Sembach K. R., et al., 2003, ApJS, 146, 165

\bibitem[Shankar et~al.(2006)]{Shankar2006}
Shankar F., Lapi A., Salucci P., De Zotti G., Danese L., 2006, ApJ, 643, 14

\bibitem[Sheth \etal(2001)]{shet01} 
Sheth R. K., Mo H. J., Tormen G., 2001, MNRAS, 323, 1

\bibitem[Shirey \etal(2001)]{Shirey2001}
Shirey R., et al., 2001, A\&A, 365, L195

\bibitem[Silk (2003)]{Silk03}
Silk J., 2003, MNRAS, 343, 249

\bibitem[Sommer-Larsen (2006)]{Sommer2006}
Sommer-Larsen J., 2006, ApJ, 644, L1

\bibitem[Strickland \& Stevens (2000)]{ss00}
Strickland D. K., Stevens I. R., 2000, MNRAS, 314, 511

\bibitem[Sutherland \& Dopita (1993)]{SD93}
Sutherland R. S., Dopita M. A., 1993, ApJS, 88, 253

\bibitem[Tang \& Wang (2005)]{Tang05}
Tang Shikui, Wang Q. D., 2005, ApJ, 628, 205


\bibitem[Takahashi \etal(2004)]{TOKM2004}
Takahashi H., Okada Y., Kokubun M., Makishima K., 2004, ApJ, 615, 242

\bibitem[Toft et~al. (2002)]{Toft02}
Toft S., Rasmussen J., Sommer-Larsen J., Pedersen K., 2002, MNRAS, 335, 799

\bibitem[Voit \etal(2003)]{voit03} 
Voit G. M., Balogh M. L., Bower R. G., Lacey C. G., Bryan G. L., 2003, ApJ, 593, 272

\bibitem[Wechsler et~al.(2002)]{Wechsler2002}
Wechsler R. H., Bullock J. S., Primack J. R., Kravtsov A. V., Dekel A., 2002, ApJ, 568, 52

\bibitem[White \& Frenk(1991)]{whit91} 
White S. D. M., Frenk C. S., 1991, ApJ, 379, 52

\bibitem[White \& Rees (1978)]{whit78}
White S. D. M., Rees M. J., 1978, MNRAS, 183, 341

\bibitem[White \& Chevalier(1983)]{WC1983}
White R. E. I., Chevalier R. A., 1983, ApJ, 275, 69

\bibitem[Wyithe \& Loeb(2003)]{wyit03} 
Wyithe J. S. B., Loeb A., 2003, ApJ, 595, 614

\bibitem[Zhao et~al.(2003a)]{Zhao03a}
Zhao D. H., Mo H. J., Jing Y. P., Borner G., 2003a, MNRAS, 339, 12

\bibitem[Zhao et~al.(2003b)]{Zhao03b}
Zhao D. H., Jing Y. P., Mo H. J., Borner G., 2003, ApJ, 597, L9


\end{thebibliography}
\end{document}